\documentclass[twocolumn,aps,superscriptaddress,showpacs,nofootinbib,floatfix]{revtex4}
\usepackage{epsfig,bm,feynmf}
\usepackage{graphics}
\usepackage{amsmath,here}

\usepackage[normalem]{ulem}  
\usepackage[dvips]{color} 

\renewcommand\sout{\bgroup \color{red} \ULdepth=-.5ex \ULset}
\begin{document}


\title{Traces of non-equilibrium effects,  initial condition,  bulk dynamics and elementary collisions  in the charm observables}


\author{Taesoo Song}\email{tsong@gsi.de}
\affiliation{GSI
Helmholtzzentrum f\"{u}r Schwerionenforschung GmbH, Planckstrasse 1,
64291 Darmstadt, Germany}

\author{Pierre Moreau}
\affiliation{Department of Physics, Duke University, Durham, North Carolina 27708, USA}
\affiliation{Institute for Theoretical Physics, Johann Wolfgang
Goethe Universit\"{a}t, Frankfurt am Main, Germany}

\author{Yingru Xu}
\affiliation{Department of Physics, Duke University, Durham, North Carolina 27708, USA}

\author{Vitalii Ozvenchuk}
\affiliation{H. Niewodnicza\'{n}ski Institute of Nuclear Physics,
Polish Academy of Sciences, 31-342 Krakow, Poland}
\affiliation{SUBATECH UMR 6457 (IMT Atlantique, Universit\'{e} de Nantes,
	IN2P3/CNRS), 4 Rue Alfred Kastler, F-44307 Nantes, France}

\author{Elena Bratkovskaya}
\affiliation{GSI
Helmholtzzentrum f\"{u}r Schwerionenforschung GmbH, Planckstrasse 1,
64291 Darmstadt, Germany}
\affiliation{Institute for Theoretical Physics, Johann Wolfgang
Goethe Universit\"{a}t, Frankfurt am Main, Germany}

\author{Jorg Aichelin}
\affiliation{SUBATECH UMR 6457 (IMT Atlantique, Universit\'{e} de Nantes,
	IN2P3/CNRS), 4 Rue Alfred Kastler, F-44307 Nantes, France}
\affiliation{FIAS, Ruth-Moufang-Strasse 1, 60438 Frankfurt am Main, Germany}

\author{Steffen A. Bass}
\affiliation{Department of Physics, Duke University, Durham, North Carolina 27708, USA}

\author{Pol Bernard Gossiaux}
\affiliation{SUBATECH UMR 6457 (IMT Atlantique, Universit\'{e} de Nantes,
	IN2P3/CNRS), 4 Rue Alfred Kastler, F-44307 Nantes, France}

\author{Marlene Nahrgang}
\affiliation{SUBATECH UMR 6457 (IMT Atlantique, Universit\'{e} de Nantes, IN2P3/CNRS), 4 Rue Alfred Kastler, F-44307 Nantes, France}


\begin{abstract}
 Heavy quarks produced in relativistic heavy-ion collisions are known to be sensitive probes of the hot and dense QCD matter they traverse. In this manuscript we study how their dynamics is affected by the nature of the bulk evolution of the QCD matter,  the initial condition of the system, and the treatment of elementary interactions between heavy quarks and the surrounding medium. For the same initial condition and the same QGP expansion scenario we discuss the consequences of the assumption of a local equilibrium by comparing the consequences for the nuclear modification factor $R_{\rm AA}$ and the elliptic flows of charm quarks, scrutinizing the different components of the final distribution of charm quarks. For this purpose we employ the  parton-hadron-string dynamics (PHSD) model, which is an off-shell microscopic transport approach, as well as the linearized-Boltzmann (LB) scheme obtained by coarse graining the PHSD bulk and assuming local equilibrium for the interactions of the charm quarks with the bulk. The $R_{\rm AA}$ of charm quarks stemming from the later LB approach is also compared to a genuine fluid dynamics evolution initiated by the coarse grained PHSD, which allows to further assess the consequences of reducing the full n-body dynamics. We then proceed to a systematic comparison of PHSD (in its LB approximation) with MC@HQ, another transport model for heavy flavors which also relies on LB approach. In particular, we investigate the consequences for the nuclear modification factor of charm quarks if we vary separately the initial heavy quark distribution function in matter, the expansion dynamics of the QGP and the elementary interactions of heavy quarks of these models. We find that the results for both models vary significantly depending on the details of the calculation. However, both models achieve very similar predictions for key heavy quark observables for certain combinations of initial condition, bulk evolution and interactions. We conclude that this ambiguity limits our ability to determine the different properties of the system based on the current set of observables.
\end{abstract}


\maketitle

\section{Introduction}

Relativistic heavy-ion collisions create extremely hot and dense  matter. At the densities and temperatures reached in these reactions  hadrons do not exist anymore and the constituents of hadrons, the quarks and gluons, form a new state of matter, a plasma of quarks and gluons (QGP).
Due to the early universe having been in a QGP state and its occurrence in dense neutron stars, the properties of the QGP are of significant interest.

One  promising probe to exhibit sensitivity to QGP properties are heavy flavor hadrons created in heavy ion collisions.
Heavy flavor particles  are distinguished from other probes because  their production can reliably be described by perturbative quantum chromodynamics (pQCD)~\cite{Cacciari:1998it,Cacciari:2001td,Vogt} and because their production and formation time is relatively short, which enables them to probe strongly interacting  matter from the early stage of heavy-ion collisions on. Heavy flavor particles are rare and calculations show that  in relativistic heavy-ion collisions only those with a low transverse momentum, $p_T$,  equilibrate with the QGP. Due to their large mass and small interaction cross section, hydrodynamics, which has been so successful in describing  the dynamics of the bulk particles of the QGP, is not applicable to heavy flavor particles. Instead, Langevin or Boltzmann equations are used~\cite{Moore:2004tg,vanHees:2005wb,Uphoff:2012gb,He:2014cla,Cao:2013ita,Gossiaux:2010yx,Das:2015ana,Song:2015sfa,Song:2015ykw,Xu:2017obm,Ke:2018tsh} to describe their time evolution.

The Langevin equation describes the time evolution of heavy flavor particles in  matter by using  drag and diffusion coefficients, which are pre-calculated as a function of the temperature and the heavy flavor momentum~\cite{Moore:2004tg,Svetitsky:1987gq}. Therefore it has to be assumed that  the environment, in which the interaction takes place, is close to local thermal equilibrium.  The  Boltzmann equation is a more general approach which does not need the assumption of a local thermal equilibrium and the interactions of heavy flavor particles with matter is expressed in terms of particle-particle interactions. Under the condition that the scattering partners of the heavy quarks are in local equilibrium, the Boltzmann equation reduces to the linearized Boltzmann (LB) equation, which is less time costly to calculate than the full Boltzmann equation and which nevertheless conserves  energy and momentum in the elementary collision between heavy quarks and their scattering partners.

However, thermalization of particles in heavy-ion collisions takes time  and therefore hydrodynamics is applicable only after the initial thermalization time, which is assumed to be around 0.6 fm/c, depending on the collision energy.
It is presently unclear how heavy quarks interact with pre-equilibrium  matter. Since in heavy-ion collisions the initial thermalization time is short compared to the lifetime of the QGP, early pre-equilibrium interactions are ignored in most studies which employ hydrodynamics.
It should also be noted that not all matter reaches a state of complete thermalization, even at freeze-out, which can be seen from the long tail of the momentum spectrum of the particles, which originates mostly from initial hard scatterings.

Some of us have recently studied the effects of non-equilibrium matter on transport coefficients of heavy quarks~\cite{Song:2019cqz}.
We showed, employing the dynamical quasi particle model (DQPM), that non-equilibrium features like an anisotropic pressure  or a deviation of the average kinetic energy or of the mass of the partons from the thermal value, modify the equilibrium transport coefficients.
In this study we continue  to investigate the effects of non-equilibrium matter on the dynamics of charm quarks in relativistic heavy-ion collisions by comparing the results with and without the assumtion of local thermal equilibrium in the Boltzmann transport approach.

For this purpose we use the parton-hadron-string dynamics (PHSD) which is based on the dynamical quasi-particle model~\cite{Cassing:2009vt}.
The PHSD has quite reasonably reproduced experimental data of relativistic heavy-ion collisions from the super proton synchrotron (SPS) to large hadron collider (LHC) energies~\cite{PHSD,PHSDrhic,Volo,Linnyk}. For the time evolution of the QGP and of the hadronic matter it employs the Kandanoff-Baym equations in which particles have a spectral function.  The  pole mass and the  width depend on the temperature and the baryon chemical potential.
This dependence is given by a hard thermal loop calculation and the strong coupling constant is fitted to the equation-of-state of strongly interacting  matter from  lattice quantum chromodynamics (lQCD).

The PHSD has been extended to the production of heavy flavor partons by using the PYTHIA event generator~\cite{Sjostrand:2006za} and the EPS 09 package for (anti)shadowing effects in heavy nuclei~\cite{Eskola:2009uj}.
Scattering cross sections of heavy quarks with off-shell parton are calculated up to leading order in the coupling constant considering dressed propagators from the DQPM~\cite{Berrehrah:2013mua,Moreau:2019vhw}.
It has been shown that the scattering cross sections reproduce the spatial diffusion coefficient of heavy quarks from lQCD calculations and the experimental data on $D$ mesons. Even more, also single electrons as well as  dileptons  are in agreement with experiment from the beam energy scan energies at RHIC to LHC energies~\cite{Song:2015sfa,Song:2015ykw,Song:2016rzw,Song:2018xca,Song:2018dvf}.

PHSD is not the only approach for heavy flavor dynamics in relativistic heavy-ion collisions. Here we compare the PHSD approach with other models, which have as well successfully described
multiple heavy flavour observables. In this comparison we keep the initial condition identical for all approaches but modify separately \\
a) the dynamics of the medium in which the heavy quarks collide (keeping the elementary interaction between the heavy quarks and the partons fixed).\\
b) the elementary interaction between the heavy quarks and the partons  (keeping the dynamics of the medium, in which the heavy quarks collide, fixed).

For the study of the influence of the bulk dynamics  we compare PHSD with causal viscous hydrodynamics which is widely used as a description of the QGP dynamics in heavy-ion collisions.
Note that hydrodynamical simulations are applicable only after an initial thermalization time and require an initial condition. PHSD can provide this initial condition such as the local energy densities, the local flow velocities or the local energy-momentum tensor at the required times. Then one can compare the dynamics of the QGP obtained from hydrodynamics with that obtained from PHSD.
It has been found, taking ensemble averages, that in the light quark sector both approaches give similar results, although in PHSD  fluctuations are much larger~\cite{Xu:2017pna}.
While in the previous study we have compared macroscopic properties of the QGP medium, such as spatial and momentum eccentricities~\cite{Xu:2017pna}, in this study we extend the comparison to heavy quark interactions with the expanding QGP  described  by hydrodynamics or by  PHSD, in order to identify how specific descriptions of the QGP dynamics affect the charm quark dynamics
in heavy-ion collisions. This comparison makes it also possible to study how the early pre-equilibrium stage modifies the observables.

Secondly we use the description of the QGP provided by the PHSD but employ  different interactions of charm quarks with the QGP.
In this way we can separate the influence of the elementary interactions from all other effects which may influence the final heavy quark spectrum.
For this comparison we use the elementary interaction advanced by the Nantes group in their MC@HQ model~\cite{Gossiaux:2009mk} to study heavy flavour production in heavy ion collisions.
This transport code for heavy flavors needs to be
supplemented with temperature and velocity fields describing the bulk
dynamics. Lately, it was then combined with another major computational model, EPOS2 \cite{Werner:2010aa} which is, as PHSD, an event generator describing the soft physics of up, down and strange quarks produced in pp, pA and AA collisions at RHIC and LHC energies. In numerous publications the results of this combined approach have been compared to experimental data. After the initial violent phase of the collision,  a quark gluon plasma (QGP) and jet-like hadrons are created. The latter do presently not interact with the QGP. The expansion of the QGP is described by hydrodynamical equations. At the transition temperature hadrons are produced utilizing the Cooper-Frye formula, and subsequent hadronic interactions are described by UrQMD \cite{Bass:1998ca,Bleicher:1999xi}.
The HQ part of the program generates heavy quarks with a FONLL distribution at the interaction points of the nucleons during the initial stage of EPOS. The heavy quarks propagate through the plasma having elastic \cite{Gossiaux:2009mk} and radiative collisions~\cite{Aichelin:2013mra,Gossiaux:2010yx}  with the plasma constituents.

When the QGP  hadronizes, the low momentum heavy quarks coalesce with a light (u,d) quark of the cell where the heavy quark is localized. Heavy quarks with high momenta hadronize by fragmentation. After fragmentation, UrQMD is used to model final hadronic interactions of the D and B mesons. Beyond the heavy flavor observables discussed here, EPOS2+MC@HQ has also been used in previous work to study  correlations between heavy quarks and antiquarks \cite{Nahrgang:2013saa}, higher order flow components \cite{Nahrgang:2014vza} and the influence of the existence of hadronic bound states beyond $T_c$ \cite{Nahrgang:2013xaa}.

This paper is organized as follows:
In section~\ref{grid} we first discuss on how to realize a coarse grained medium in  PHSD.
Section~\ref{assumption} shows how the assumption of local thermal equilibrium affects charm quark interactions in heavy-ion collisions by using a linearized Boltzmann approach.
Section~\ref{yingru} is devoted to the comparison of charm dynamics in the PHSD with that in 3+1 dimensional viscous hydrodynamics initialized by the PHSD and also discusses the effects of pre-equilibrium interactions on charm in heavy-ion collisions.
We then study the effects of different initial conditions and heavy quark - light parton interactions on common observables in section~\ref{nantes}, comparing results from PHSD and MC@HQ. Finally, a summary is given in section~\ref{summary}.

\section{Coarse-graining the PHSD medium}\label{grid}

In order to study the non-equilibrium effect on charm and to compare with other models, the coarse-graining of the PHSD medium is necessary.
For example, one can introduce local thermal equilibrium to the coarse graining of the PHSD and compare with the charm from the original PHSD, and the difference will be the non-equilibrium effect on charm in heavy-ion collisions, which is described in section~\ref{assumption} in details.
It also enables to compare between models, because many models which study charm in heavy-ion collisions assume local thermal equilibrium.

To calculate local thermal quantities in PHSD such as the energy density or the energy-momentum tensor  we introduce a grid. During the expansion
one projects all particles onto the corresponding grid and calculates these quantities cell by cell.  In relativistic heavy-ion collisions this coarse graining procedure needs special care, due to the relativistic expansion of the QGP medium along the beam axis.
In  PHSD,  the grid size is fixed to 1 fm in the x and y directions respectively.
Since the matter expands almost with the speed of light in z-direction, the grid size in z-direction is designed to grow with time.
Before the two nuclei pass through each other, the grid size along z-direction and the time step are, respectively, given by
\begin{eqnarray}
dz=\frac{1}{\gamma_{cm}},~~~dt=\frac{dz}{2}
\label{dzdt-ini}
\end{eqnarray}
where
\begin{eqnarray}
\gamma^{cm}=\frac{1}{2}\bigg(\frac{E^{\rm projectile}}{M^{\rm projectile}}+ \frac{E^{\rm target}}{M^{\rm target}}\bigg).
\end{eqnarray}
We note that $dt$ is taken to be smaller than $dz$ in order not to violate causality.
 In each nucleus rest frame $dz$ in Eq.~(\ref{dzdt-ini}) is 1 fm, as $dx$ and $dy$.
After the passage of the two nuclei, $dz$ grows linearly with time as
\begin{eqnarray}
dz\approx \frac{1}{N_z}(t-t^*)+\frac{1}{\gamma_{cm}}.
\label{dzdt-expan}
\end{eqnarray}
$t^*$ is the approximate time which two nuclei need to pass each other and $N_z$ is the number of grid cells in +(-) z-direction. Eq.~(\ref{dzdt-expan}) implies that the grid size in +(-) z-direction corresponds to the elapsed time after $t^*$:
\begin{eqnarray}
z_{max}=N_z\times dz\approx t-t^*.
\end{eqnarray}

One can also use grid cells in the  ($\tau, x, y, \eta$) frame where $\tau$ is the  longitudinal proper time and $\eta$ the spatial rapidity,
\begin{eqnarray}
\tau&=&\sqrt{t^2-z^2},\label{def-tau}\\
\eta&=&\frac{1}{2}\ln\bigg(\frac{t+z}{t-z}\bigg).\label{def-eta}
\end{eqnarray}

This coordinate system is very convenient to describe matter which is boost-invariant as approximately realized in relativistic heavy-ion collisions.  Therefore hydrodynamic simulations and many fireball models often use this coordinate system. It is, however,  a bit tricky to use this coordinate system in Boltzmann-type transport models, because particle position and momentum should then be updated based on $d\tau$, not on $dt$ whereas the update in the PHSD transport equations is done in $dt$.

\begin{figure}[h!]
\centerline{
\includegraphics[width=8.6 cm]{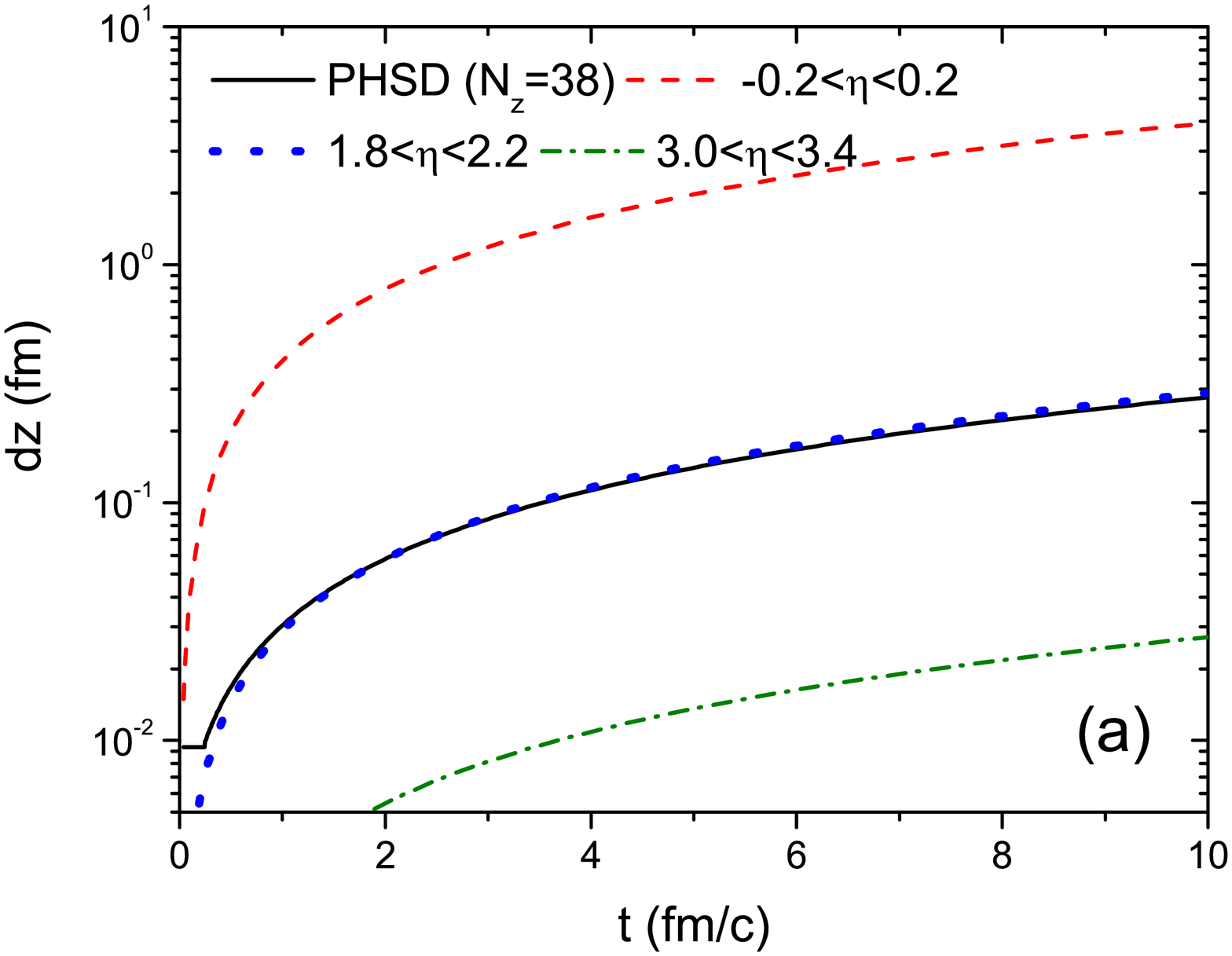}}
\centerline{
\includegraphics[width=8.6 cm]{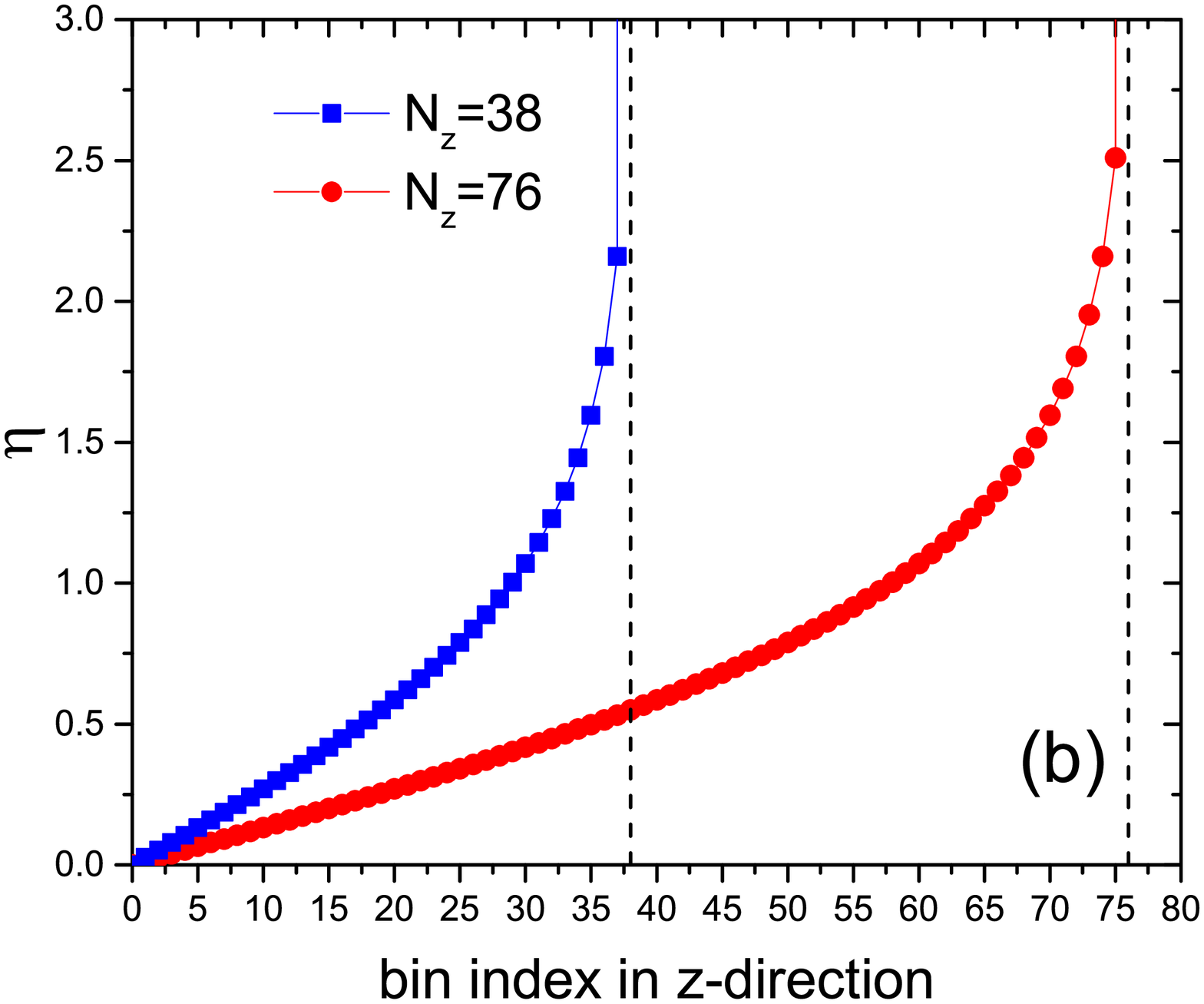}}
\caption{(a) Longitudinal cell size in PHSD compared with those from constant $\eta$ as functions of time and (b) $\eta$ corresponding to each cell boundary in (t,z) coordinate system for $N_z=$ 38 and 76.}
\label{dz}
\end{figure}
In figure~\ref{dz} (a) we see the difference between the grid in the Cartesian coordinate system ($t, z$) and that in ($\tau, \eta$).
The black line is the cell size in z-direction as a function of time given by Eq.~(\ref{dzdt-expan}). It does not depend on the position of the cell.
On the other hand, the dashed, dotted, and dash-dotted lines are calculated for constant $d\eta=0.4$ bins
as function of time.
$dz$ at a fixed $t$ is given as
\begin{eqnarray}
dz&\approx& \frac{1}{N_z}\times t, \quad  \mathrm{ for\ a \ (t,z) grid}\\
dz&=&{\rm sech}^2(\eta)d\eta \times t  \quad  \mathrm{ for\ a\ (\tau,\eta) grid},
\end{eqnarray}
where the first $dz$ does not depend on $z$ or $\eta$, while the second one
depends on $\eta$ and $d\eta$ and is smaller for a larger $\eta$ because of the factor, ${\rm sech}^2(\eta)$.
Since $N_z$ is 38 in PHSD, $dz$ in PHSD is similar to $dz$ for $1.8<\eta<2.2$, as shown in figure~\ref{dz}.
We can conclude that the coordinate system ($t, z$) has a better resolution at mid-rapidity while that of ($\tau, \eta$) is better at forward and backward rapidities, if the same number of grid cells is used.

Figure~\ref{dz} (b) shows $\eta$ corresponding to each cell boundary in the (t,z) coordinate system for $N_z=$ 38 and 76.
One can see that central cells in (t,z) coordinate system correspond to very small $d\eta$, and corresponding $d\eta$ increases with larger cell index.
In the case of $N_z=$ 38 the second last cell covers 1.8 $< \eta <$ 2.2 and the last cell  2.2 $< \eta < \infty$.
Increasing $N_z$ by a factor of two,  $N_z=$ 76, the last cell covers still a large range 2.5 $< \eta < \infty$.

In the next section we use both coordinate systems to study the charm dynamics in relativistic heavy-ion collisions with  PHSD.
It is straightforward to calculate the local energy-momentum tensor or the energy density in the coordinate system ($t, z$) while the calculations in the coordinate system ($\tau, \eta$) needs a brief description.

In the PHSD approach particles are updated with a constant time step  $\Delta t$.  Hence we know positions and momenta of all particles at times   $t_i=t_0+i\cdot \Delta t$ where $i$ is a positive integer number.  We can calculate $\eta$ at $t_i$ from Eq.~(\ref{def-eta}) and also $dz$ corresponding to a constant $\Delta \eta$:
\begin{eqnarray}
dz_{(i,j)}=z_{(i,j+1)}-z_{(i,j)}=t_i\{\tanh(\eta_{j+1})- \tanh(\eta_{j})\},\nonumber\\
\end{eqnarray}
where $i$ is the time index and $j$ is the index of the spatial rapidity with $\Delta \eta=\eta_{j+1}-\eta_{j}$.
As a next step, the energy-momentum tensor of the cell, whose boundaries are $z_{(i,j)}$ and $z_{(i,j+1)}$, is calculated, and the energy density and flow velocity are obtained by diagonalization~\cite{Xu:2017pna}.
We assume that the calculated energy density and the flow velocity is located at the center of the cell,
\begin{eqnarray}
(t,z)=\bigg(t_i,~\frac{z_{(i,j+1)}-z_{(i,j)}}{2}\bigg),
\label{correspondence}
\end{eqnarray}
and the information is transferred into a cell in ($\tau, \eta$) coordinate system by using Eqs.~(\ref{def-tau}) and (\ref{def-eta}).
In this case $dz$ and $d\eta$ are in one-to-one correspondence while $dt$ and $d\tau$ are not.
Since the size of $dt$ in  PHSD is small, several cells in $t$-direction correspond to one cell in the $(\tau,\eta)$ grid.
We solve this problem by taking averages over the energy densities and the flow velocities of  several cells for the one cell in $(\tau,\eta)$ grid.

\section{assumption on local thermal equilibrium}\label{assumption}

In the grid defined above, the energy-momentum tensor is calculated for each cell. Then energy density, pressure, and flow velocity are obtained by diagonalizing the energy-momentum tensor~\cite{Xu:2017pna}.
Since the matter produced in heavy-ion collisions is not necessarily in thermal equilibrium, the pressure is,   especially in the early stage, not isotropic.
Compared to the isotropic pressure of a thermalized QGP at the same energy density, the transverse pressure in PHSD is initially small and increases with time, until it converges
to the isotropic pressure before $\tau=1$ fm/c in central Au+Au collisions at $\sqrt{s_{\rm NN}}=$200 GeV~\cite{Xu:2017pna}.
Extracting the  longitudinal pressure is technically difficult, since it depends on the longitudinal size of cell.
If the longitudinal size of cell is chosen too large, the longitudinal flow will contribute to the longitudinal pressure.
On the other hand, a too small longitudinal size will provoke large fluctuations due to the small average number of particles in the cell, and the calculation of the longitudinal pressure becomes very difficult.

The parton mass and the strong coupling in PHSD depend on the  temperature. If the system is not in complete equilibrium we calculate
the temperature and a chemical potential with help of the equation of state (which is the lattice equation of state)
by using the local energy density and baryon density as input.

Heavy quarks produced in heavy-ion collisions interact with  the QGP composed of quarks and gluons.
Quarks with high transverse momentum lose a considerable amount of energy while quarks with low transverse momentum gain energy due to the collective flow.
Interactions are described in PHSD by the scattering of heavy quarks with individual partons.
This microscopic approach is time-consuming, since the energy-momentum and the position of each parton is updated at each time step during their propagation through the medium and possible collision partners need to be identified during each time step as well.

A simpler, alternative, method is the linearized Boltzmann (LB) approach, where the light partons from the QGP are assumed to be so close to thermal equilibrium that small contributions from non-equilibrium effects can be ignored in the Boltzmann collision integral:
\begin{eqnarray}
f(k)=f_{\rm eq.}(k,T)+\delta(k)\approx f_{\rm eq.}(k,T).
\label{linearizedB}
\end{eqnarray}
$f(k)$ is the real momentum distribution of the  partons, $f_{\rm eq.}(k,T)$ is the thermal distribution at a given temperature $T$ and $\delta(k)$ is a small deviation from the equilibrium distribution.
We note that the above approximation applies for the distribution of the QGP partons  but not to that of heavy quarks which may be far away from equilibrium with the QGP particles.
Assuming Eq.~(\ref{linearizedB}), one can calculate the interaction rate of heavy quarks:
\begin{eqnarray}
\Gamma = \frac{1}{2E_p}\sum_{i=q,\bar{q},g}\int \frac{d^3k}{(2\pi)^3 2E}f_i(k,T)\int \frac{d^3k^\prime}{(2\pi)^3 2E^\prime}\nonumber\\
\times\int \frac{d^3p^\prime}{(2\pi)^3 2E_p^\prime}~(2\pi)^4\delta^{(4)}(p+k-p^\prime-k^\prime)\frac{|M_{ic}|^2}{\gamma_c},
\label{interaction-rate}
\end{eqnarray}
with $(E_p,~p),~(E,~k)$ being the energy-momenta of the heavy quark $c$ and of the scattering partner $i$ before scattering and $(E_p^\prime, ~p^\prime),~(E^\prime, ~k^\prime)$ being those after scattering,  respectively. $M_{ic}$, $\gamma_c$ and $f_i(k,T)$ are the  scattering amplitude, the degeneracy factor of heavy quarks, and the distribution function of the scattering partner $i$ at the temperature $T$, respectively.

In DQPM, which is employed in  PHSD, partons are described by a spectral function~\cite{Linnyk:2015rco}:
\begin{align}
	\rho(k_0,{\bf k}) & = \frac{\gamma}{\tilde{E}}
	\left(\frac{1}{(k_0-\tilde{E})^2+\gamma^{2}}
	-\frac{1}{(k_0+\tilde{E})^2+\gamma^{2}}\right) \nonumber \\
	& \equiv \frac{4k_0\gamma}{\left( k_0^2 - \mathbf{k}^2 - M^2 \right)^2 + 4\gamma^2 k_0^2},
	\label{spectral_function}
\end{align}
where $\tilde{E}^2({\bf k})={\bf k}^2+M^{2}-\gamma^{2}$ with $\gamma$ and $M$ being the spectral width and the pole mass, respectively. Both are functions of the temperature and the baryon chemical potential. Considering the normalization of the spectral function,
\begin{eqnarray}
\int_{-\infty}^{\infty}\frac{dk_0}{2\pi}k_0\rho(k_0,{\bf k})=\int_0^{\infty}\frac{dk_0}{2\pi}2k_0\rho(k_0,{\bf k})=1,
\end{eqnarray}
the interaction rate in Eq.~(\ref{interaction-rate}) is covariantly expressed by
\begin{eqnarray}
\Gamma = \frac{1}{2E_p}\sum_{i=q,\bar{q},g}\int \frac{d^4k}{(2\pi)^4}f_i(k,T)\rho_i(k,T)\int \frac{d^4k^\prime}{(2\pi)^4}\rho_i(k^\prime,T)\nonumber\\
\times\int \frac{d^3p^\prime}{(2\pi)^3 2E_p^\prime}~(2\pi)^4\delta^{(4)}(p+k-p^\prime-k^\prime)\frac{|M_{ic}|^2}{\gamma_c},~~~\nonumber\\
\label{covariant}
\end{eqnarray}
where the charm spectral function is substituted by a delta function,
\begin{eqnarray}
\rho(E_p^\prime,p^\prime)\rightarrow 2\pi\delta^+(p^{\prime 2}-m_c^2).
\end{eqnarray}
$m_c$ is the heavy quark mass. 
 In this study a nonrelativistic approximation is taken to  Eq. (\ref{spectral_function}), and the Breit-Wigner spectral function $\rho(m)$,
\begin{equation}
\frac{k_0}{\pi}\rho(k_0,{\bf k})\rightarrow \rho^{BW}(m) = \frac{2}{\pi} \frac{2m^2 \gamma}{(m^2-M^2)^2+(2m \gamma)^2},
\label{Breit-Wigner}
\end{equation}
is employed. The normalization is satisfied as  $\int_{0}^{\infty} dm\ \rho^{BW}(m) = 1$.

The LB approach is realized in PHSD as follows: Each heavy quark is located in a cell which has a temperature and a flow velocity.
The heavy quark  is then boosted to the cell-rest-frame (i.e. the heat-bath frame) and one obtains  the heavy quark velocity in the heat-bath frame.
The interaction rate as a function of the temperature and the heavy quark velocity in the heat-bath frame  is calculated with help of Eq.~(\ref{covariant}).
Since one needs the interaction rate in the simulation frame, it is boosted back with the opposite sign of flow velocity.
This  is simply realized by substituting $E_p$ in the denominator of Eq.~(\ref{covariant}) by the heavy quark energy in the simulation frame. The other part of the equation is Lorentz-invariant.


From the interaction rate in the simulation frame, one can decide, by using a Monte-Carlo approach, whether a heavy quark scattering takes place in the following time step or not.
One draws a random number. If it is smaller than $\Gamma_{\rm simulation}\Delta t$, with $\Delta t$ being the size of the time step in the simulation, the heavy quark will scatter.
Since $\Gamma_{\rm simulation}\Delta t$ is supposed to be less than 1, one needs to ensure that $\Delta t$ is sufficiently small.

When a collision takes place, the details of the scattering are again determined  using Monte-Carlo methods in the cell rest system.
This approach allows us to use the same collision term as it is used in PHSD for non-equilibrium matter.
%



Using the above formalism  we can now  compare three distinct scenarios:\\
1) The charm quarks  interact with gluons and light (anti)quarks   whose time evolution is given by the PHSD equations. In this approach one calculates the trajectories of all particles and therefore one does not assume that the expanding system is in local equilibrium.\\
2) The charm quarks  interact with gluons and  light (anti)quarks which are propagated as in 1) but it is assumed that they are in local equilibrium. The thermodynamical quantities are determined from the energy density and the flow velocity of the PHSD particles  in the  cell in which the heavy quark is localized,  using the equation of state. The   scattering partners of the heavy quarks are taken from the thermal parton distribution.\\
3) The charm quarks  interact with  gluons and light (anti)quarks which are assumed to be in a local equilibrium.  As in 2) the thermodynamical quantities are determined
from the properties of the cell in which the heavy quark is localized. However, these quantities are now provided by a hydrodynamical calculation of the expanding medium utilizing initial conditions generated by PHSD.\\

The elementary interaction between the heavy quarks and the gluons or light (anti)quarks are identical in all three cases and, as discussed above, are treated numerically  in an identical way. Therefore, the influence of local non-equilibirum effects can  directly be observed by comparing scenarios 1) and 2). The difference between the global expansion scenario of PHSD and a hydrodynamical expansion can be obtained by comparing 2) and 3).

\begin{figure}[h!]
\centerline{
\includegraphics[width=8.6 cm]{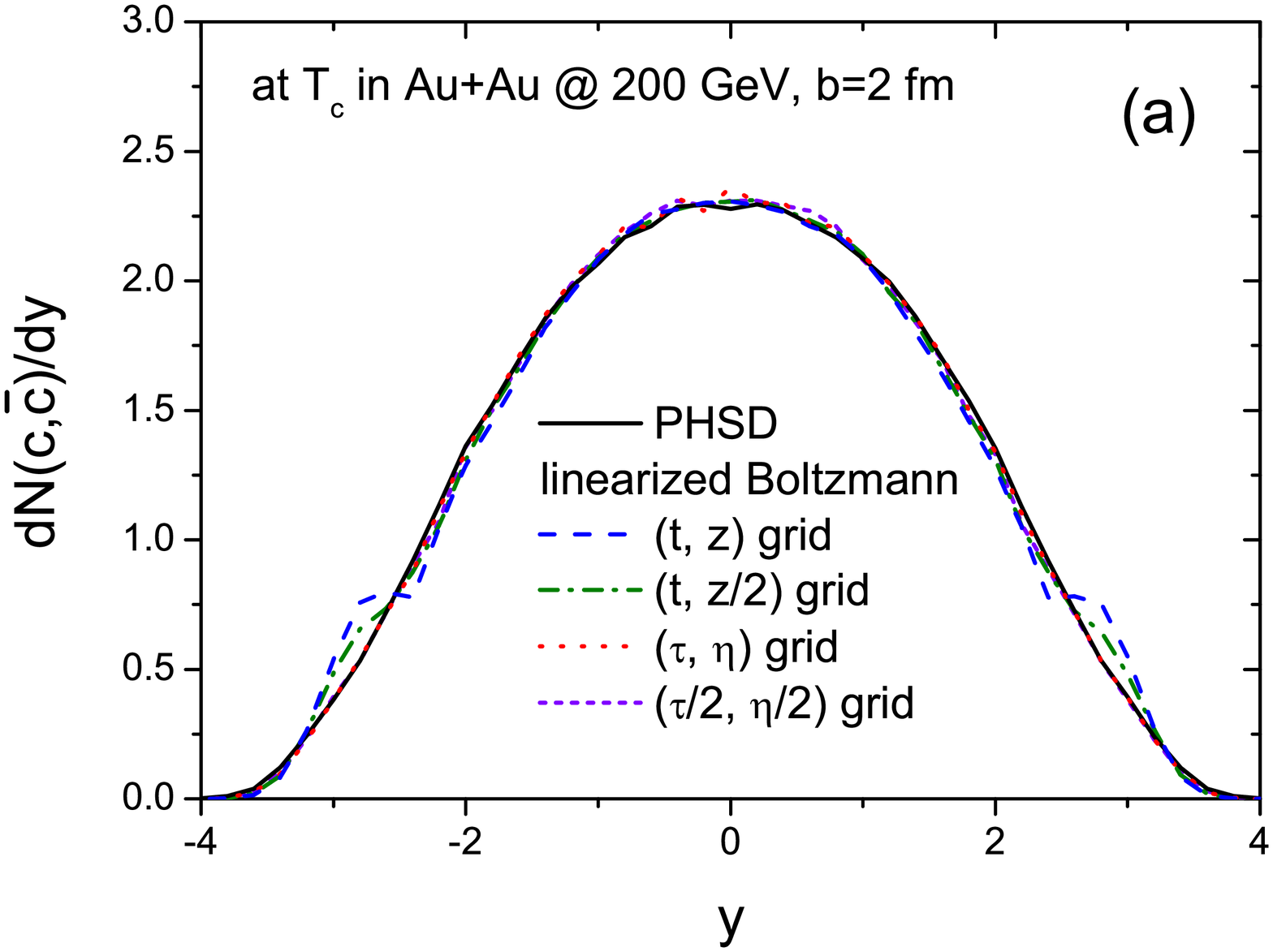}}
\centerline{
\includegraphics[width=8.6 cm]{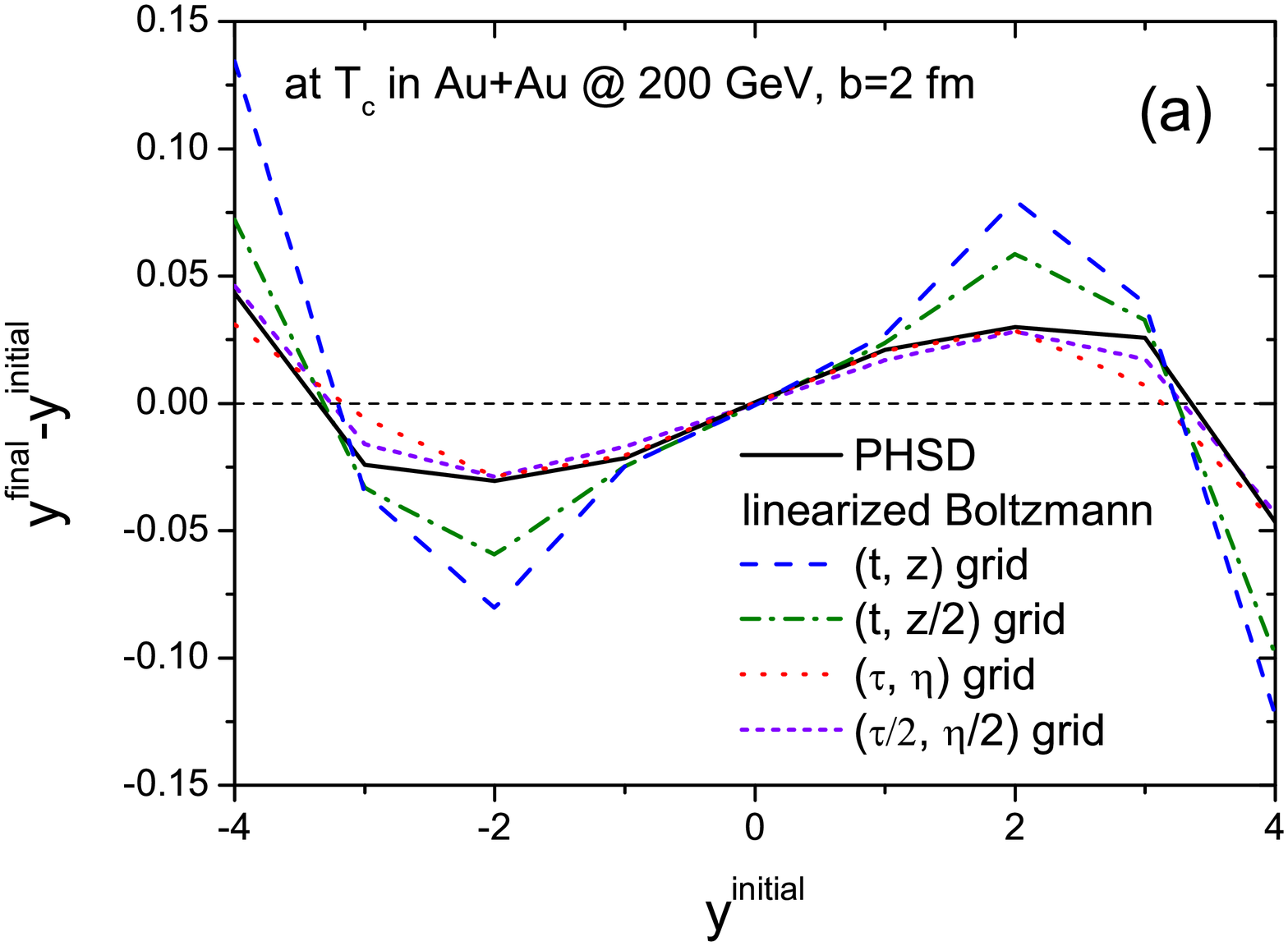}}
\caption{(a) Final rapidity distribution and (b) rapidity change  (rapidity at $T_c$ subtracted by the initial rapidity) of a (anti)charm quark as a function of the initial rapidity  in Au+Au collisions at $\sqrt{s_{\rm NN}}=$ 200 GeV and $b=$ 2 fm from PHSD and from the linearized Boltzmann approach with grids in different coordinate systems and of different sizes.}
\label{tc}
\end{figure}

Figure~\ref{tc} shows the rapidity distribution and the rapidity change (rapidity at $T_c$ subtracted by the initial rapidity) of (anti)charm quarks as a function of the initial rapidity in Au+Au collisions at $\sqrt{s_{\rm NN}}=$ 200 GeV at an impact parameter $b=$ 2 fm from PHSD and from the LB approach with grids defined in different reference frames and of different sizes.
As explained in the previous section, we can define grids in both, $(t,x,y,z)$ and $(\tau,x,y,\eta)$,  reference frames. Here we use cell sizes of $N_z=38$ and $N_z=76$ in Eq.~(\ref{dzdt-expan}) for the former case, which are denoted respectively by $(t,z)$ and $(t,z/2)$, and of $(d\tau=0.2,~d\eta=0.4)$ and $(d\tau=0.1,~d\eta=0.2)$ for the latter case, which are denoted respectively by $(\tau,\eta)$ and $(\tau/2,\eta/2)$ in the figure.

We can see in the upper panel of figure~\ref{tc} that for all 4 grids the charm rapidity distribution is almost the same near mid-rapidity but it has humps at $2<|y|<3$ in the LB approach using a grid in the $(t,z)$ coordinate system.
The reason can be seen from  the lower panel of the figure, which shows the average rapidity change of charm and anticharm quarks during the QGP phase.
Both,  the PHSD and the LB approach, show that charm quarks, which have initially a forward rapidity, are accelerated forward and those which have initially a negative rapidity are accelerated backwards. In other words, $R_{\rm AA}(y)$, the ratio of the rapidity distribution of charm quarks in heavy ion collisions versus proton-proton collisions properly scaled by the number of binary collisions, becomes larger than one at forward and backward rapidities after the time-evolution of the QGP matter.
This difference in the rapidity change for the different grids is most pronounced around
$|y|\approx 2$.

The rapidity change is largest for the LB approach with grids in the $(t,z)$ coordinate system, while using a grid in $(\tau,\eta)$  the results are similar to those in  PHSD which does not assume equilibrium. Even if the cell size is reduced to $(t,z/2)$, rapidity changes at around $|y|=2$ are still about twice as large as those observed in PHSD.
We attribute this behavior of the grid in the $(t,z)$ reference frame to its  poor resolution at forward and backward rapidities, as shown in figure~\ref{dz} (b).
Therefore it is highly recommended to use grid in $(\tau,\eta)$ reference frame to study forward and backward rapidities.

\subsection{mid-rapidity}

\begin{figure}[h!]
\centerline{
\includegraphics[width=8.6 cm]{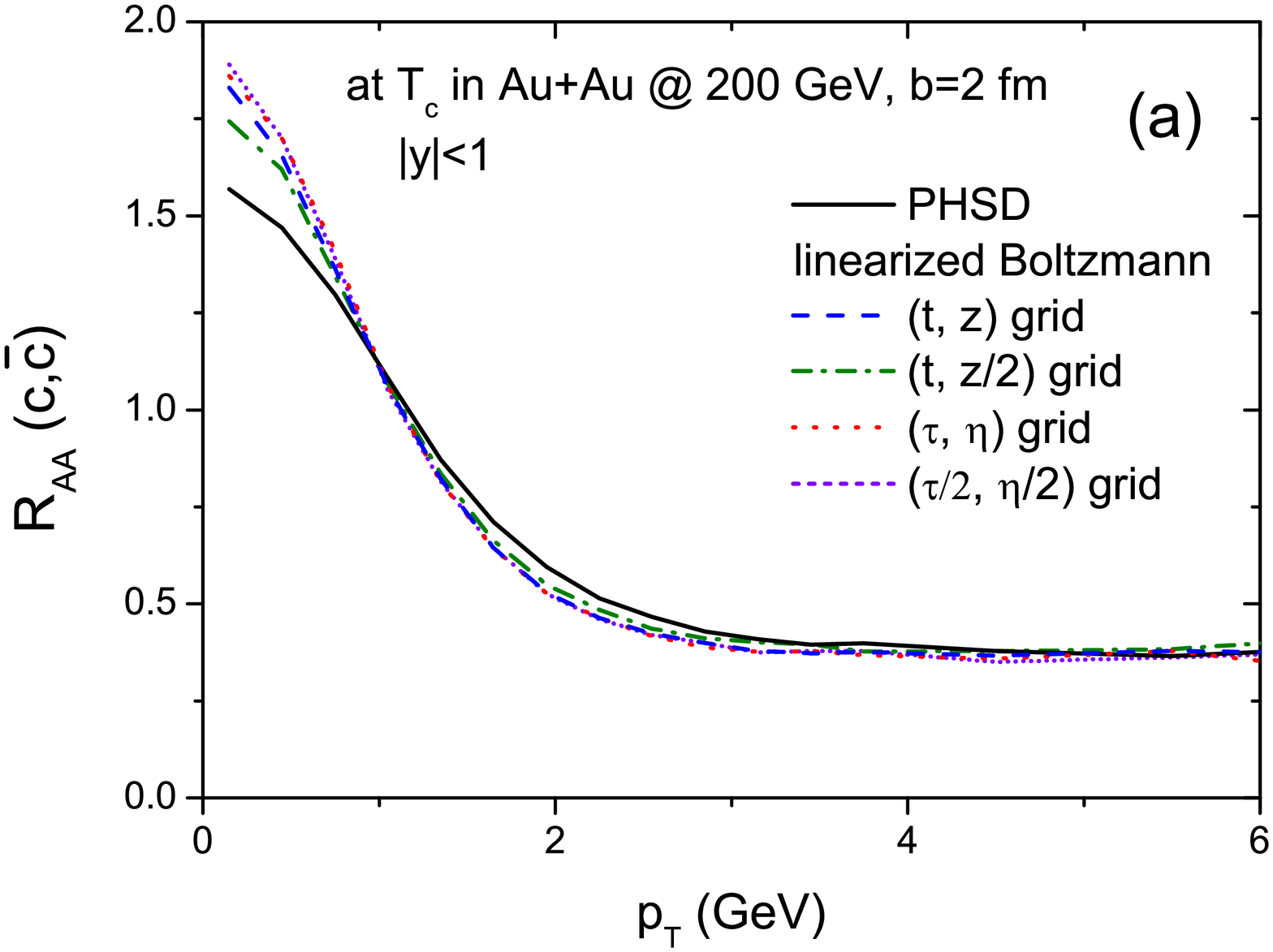}}
\centerline{
\includegraphics[width=8.6 cm]{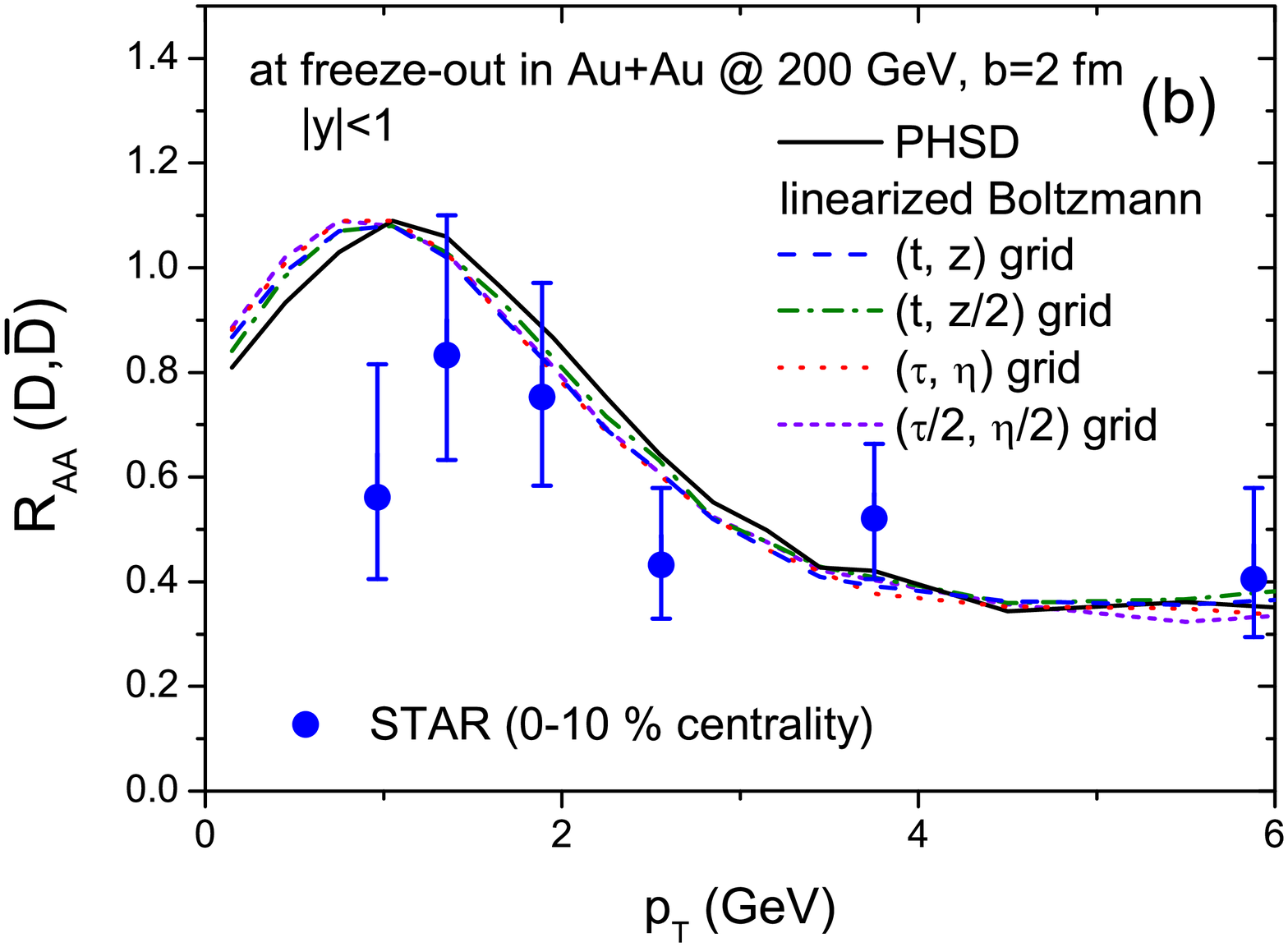}}
\centerline{
\includegraphics[width=8.6 cm]{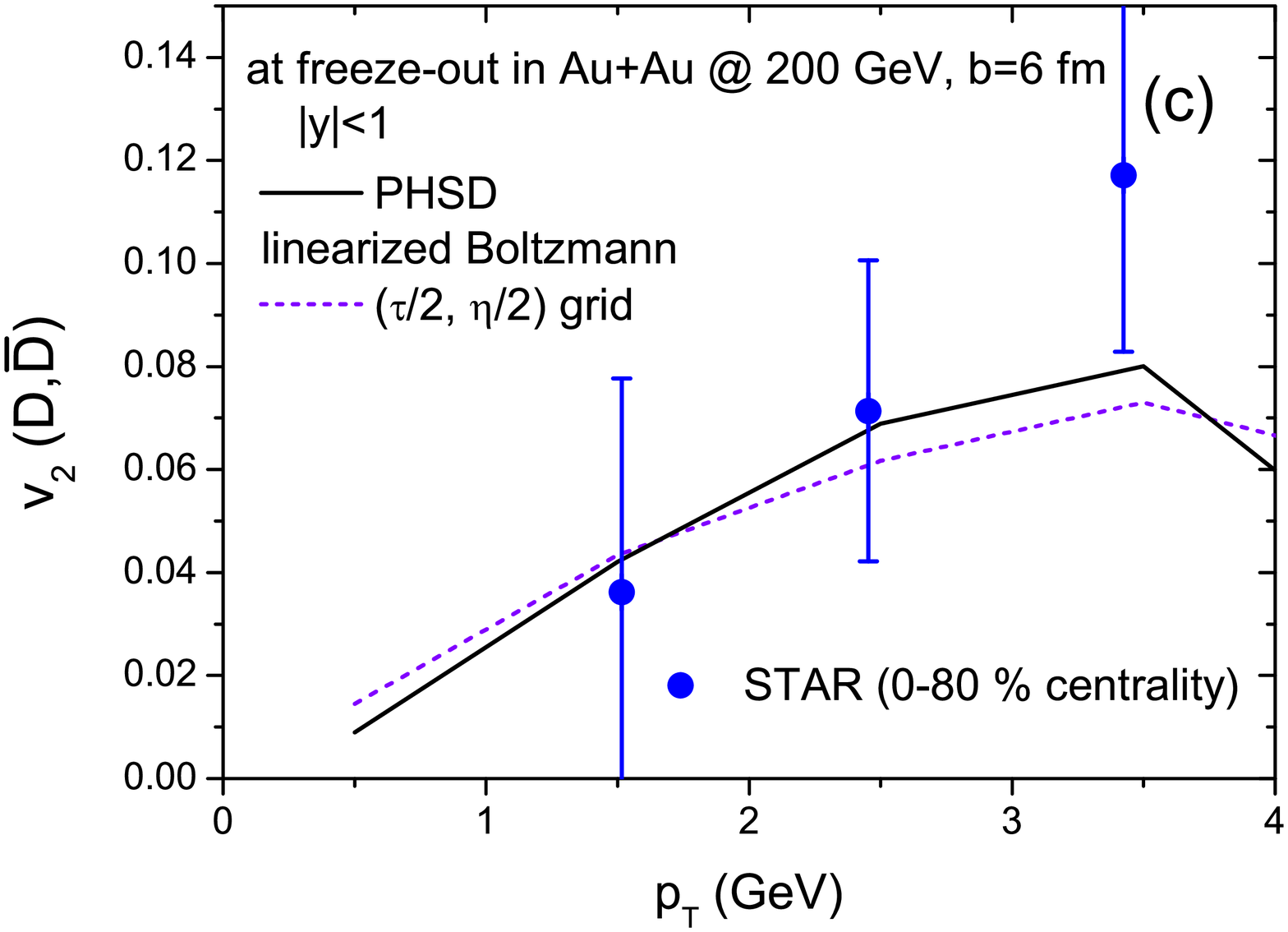}}
\caption{(a) $R_{\rm AA}$ of (anti)charm quarks at $T_c$ before hadronization, (b) $R_{\rm AA} $ and (c) the elliptic flow, $v_2$, of $D(\bar{D})$ mesons at freeze-out at mid-rapidity ($|y|<1$) in Au+Au collisions at $\sqrt{s_{\rm NN}}=$ 200 GeV. The impact parameter is $b=$ 2 fm for (a) and (b) and $b=$ 6 fm for (c) . We display
results from PHSD and from the linearized Boltzmann approach with a couple of different grids. We note that the impact parameters do not exactly correspond to the centralities of the experimental data from the STAR Collaboration~\cite{Adamczyk:2014uip,Lomnitz:2016rpz}.}
\label{raacD}
\end{figure}

We now discuss the effects of  non-equilibrium vs. equilibrium medium evolution on charm quarks at mid-rapidity.
Figure~\ref{raacD} shows $R_{\rm AA}$ of (anti)charm quarks at $T_c$ before hadronization as well as $R_{\rm AA}$ and  the elliptic flow, $v_2$ , of $D(\bar{D})$ mesons at freeze-out at mid-rapidity ($|y|<1$) in Au+Au collisions at $\sqrt{s_{\rm NN}}=$ 200 GeV. We  compare these results  with the experimental data from the STAR collaboration~\cite{Adamczyk:2014uip,Lomnitz:2016rpz}, although our impact parameter does not exactly correspond to the centrality of the experimental data.
As expected from figure~\ref{tc}, local non-equilibrium effects of the matter do not have a significant consequences for heavy flavor observables, at least for Au+Au collisions
at the top RHIC energy.
In the LB approach, for all coordinate systems and all grid sizes, $R_{\rm AA}$ of the charm quarks is larger at low transverse momentum ($p_{\rm T}<1$ GeV) and a bit smaller around $p_{\rm T}=2$ GeV, as compared to $R_{\rm AA}$ from the PHSD.

After the charm quark is hadronized into a $D$ meson, it interacts with the hadron gas until freeze-out.
We do not use the LB approach for $D$ meson scattering in the hadron gas phase but use the geometric method of  PHSD in which the hadrons interact by cross sections without assuming that they are in equilibrium. In other words, hadronization and hadronic interactions are the same in both cases.
Usually hadronization and hadronic interactions shift the maximum of the $R_{\rm AA}$ curve to a higher transverse momentum, due to coalescence with light (anti)quarks, which is the dominant hadronization mechanism at low $p_T$, and which enhances the transverse momentum of the $D$ mesons and also the radial flow becomes stronger with time. This we observe comparing $R_{\rm AA}$ in the upper panel of figure~\ref{raacD} with the  $R_{\rm AA}$   in the middle panel.
Differences between $R_{\rm AA}$ from the PHSD and that from the linearized Boltzmann approach are, however, much smaller than the experimental errors. The same is true for $v_2$.
The differences for the elliptic flow of $D$ mesons are small in comparison with the large experimental errors, as shown in the lower panel of figure~\ref{raacD}.
As we shall see, however,  the above results do not indicate that the charm interactions are similar on a microscopic level.

\begin{figure*}[h!]
\centerline{
\includegraphics[width=8.6 cm]{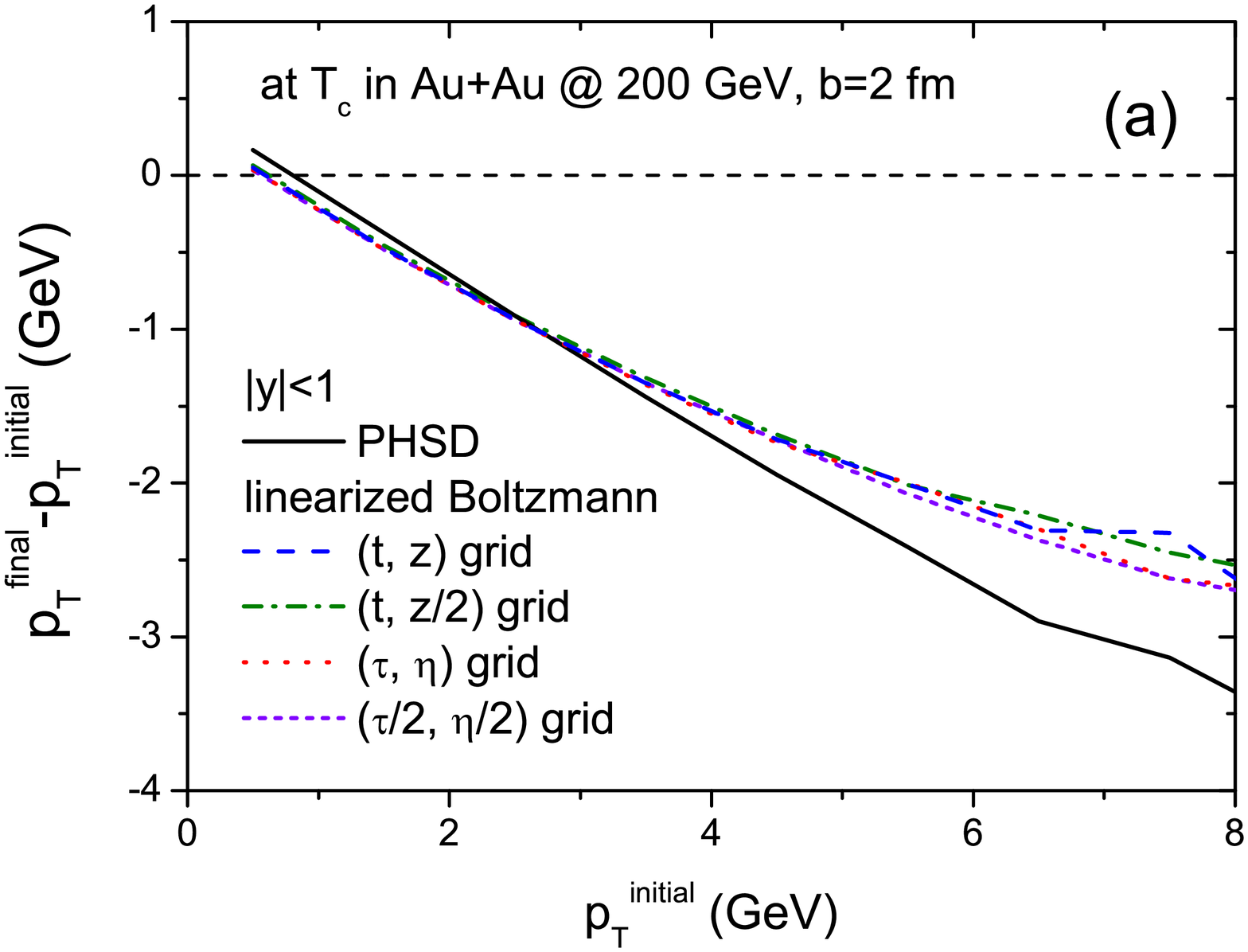}
\includegraphics[width=8.6 cm]{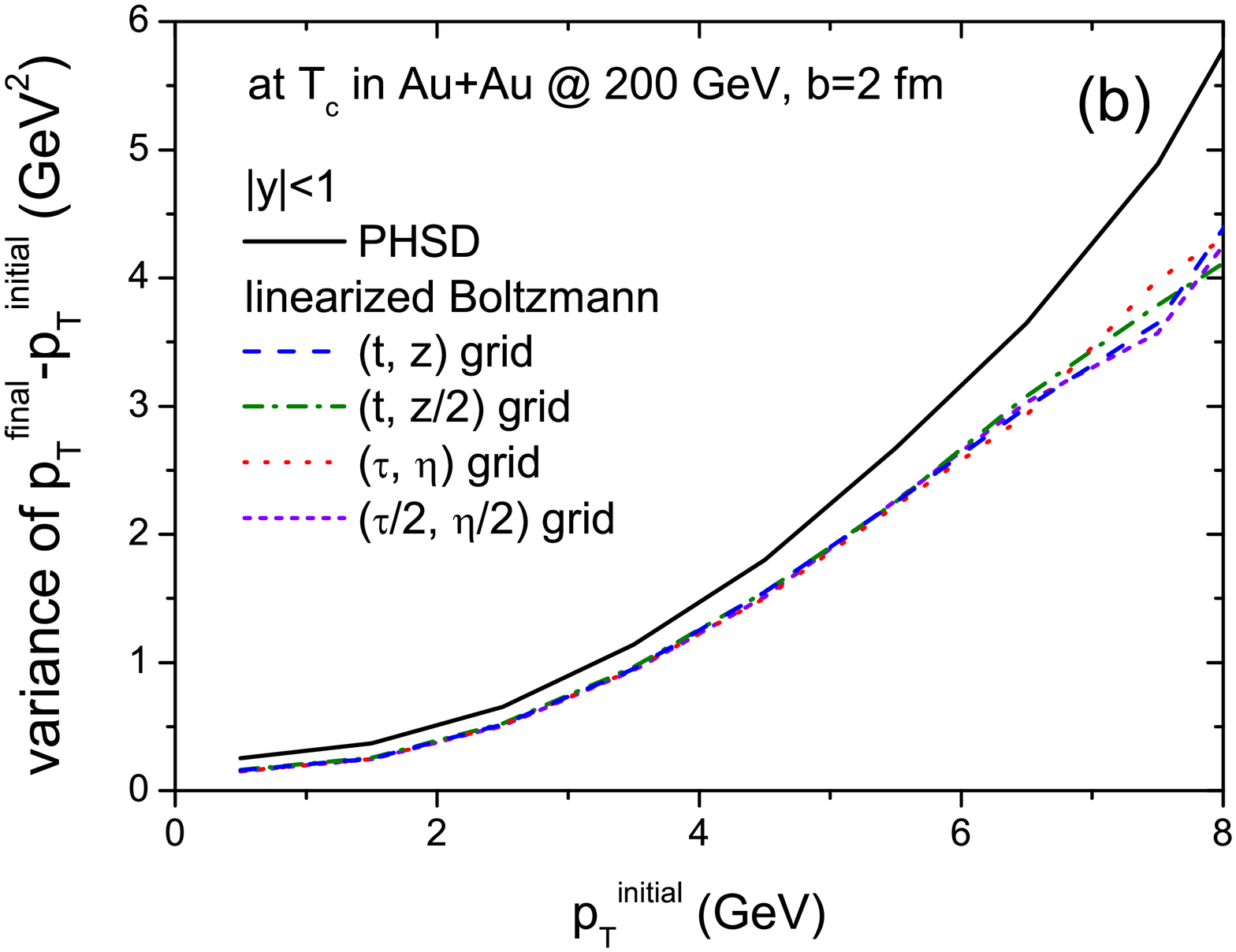}}
\centerline{
\includegraphics[width=8.6 cm]{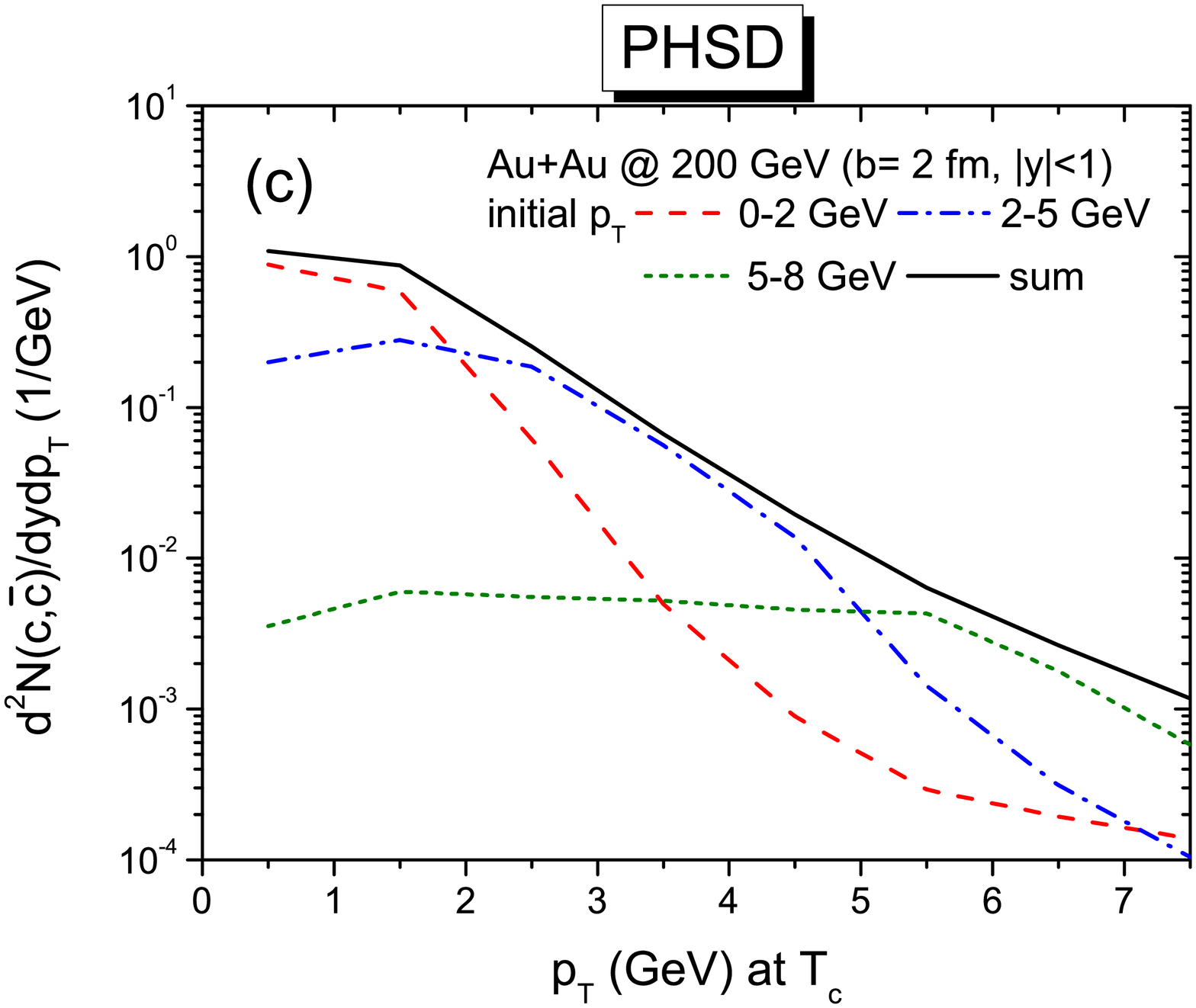}
\includegraphics[width=8.6 cm]{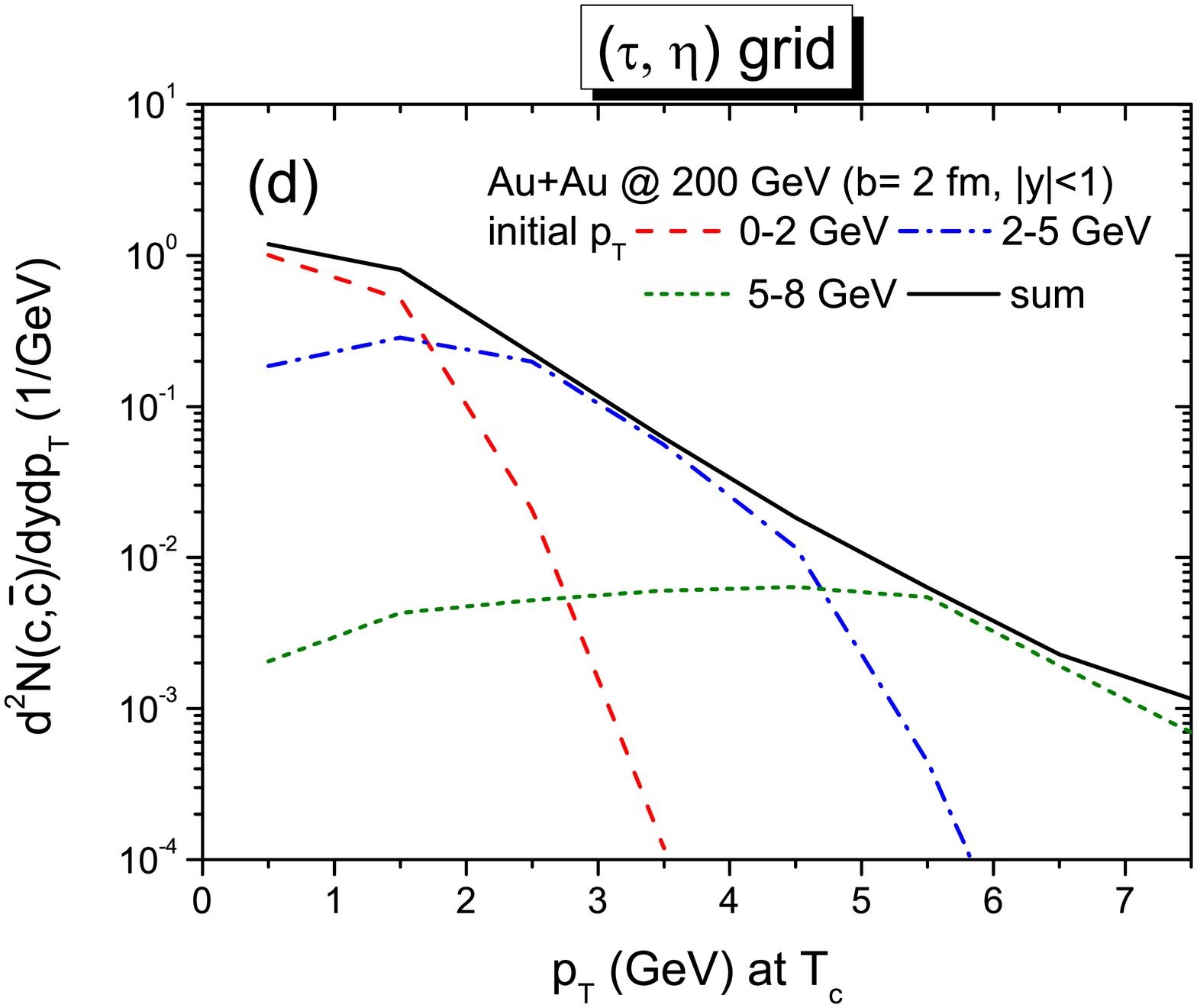}}
\caption{Top: transverse momentum change (left) and  variance of the transverse momentum (right) of (anti)charm quarks with $|y|<1$ for different grids in the LB approach and in PHSD. Bottom: Final $p_T$ distribution of charm quarks separated for different initial heavy quark $p_T$ momenta [ 0-2, 2-5, 5-8] GeV. On the left we display the results of PHSD on the right those for the LB approach. We investigate  central Au+Au collisions at $\sqrt{s_{\rm NN}}=$ 200 GeV.}
\label{microscopic}
\end{figure*}

The two upper panels of figure~\ref{microscopic} show the transverse momentum change (left)  and the variance (right) of mid-rapidity (anti)charm quarks in a QGP produced in central Au+Au collisions at $\sqrt{s_{\rm NN}}=$ 200 GeV.
Even though the $R_{\rm AA}$ and the  $v_2$ of charm quarks are similar in  PHSD and in the LB approach, the change of the transverse momentum and the variance are different.
Irrespective of the reference frame used for the grid and the grid size, in PHSD charm quarks with initially small transverse momentum gain  more $p_T$  and those which have initially a large transverse momentum lose more $p_T$, compared to  the LB approach, which assumes local thermal equilibrium.
The variance of the transverse momentum change is always larger in  PHSD than for the LB approach.
In other words, the drag of charm quarks in $p_T$ direction and its variance  is larger in PHSD
than in the LB approach. Naively one would think that a larger drag coefficient causes a larger suppression of charm quarks at high momentum.
Figure~\ref{raacD} shows, however, that $R_{\rm AA}$ of charm quarks is almost the same in  PHSD and in the LB approach. The reason for this
can be found in the lower panels of figure~\ref{microscopic}.

The two lower panels display the final  transverse momentum distributions of charm quarks at $T_c$ in central Au+Au collisions at $\sqrt{s_{\rm NN}}=$ 200 GeV from the PHSD and from the LB approach with a grid in the $(\tau, \eta)$ reference frame ($\Delta\tau=0.2$ fm/c, $\Delta\eta=0.4$).
The black solid line includes all contributions regardless of the  initial transverse momentum.
The  red dashed line, the blue dot-dashed line, and the green short dashed lines are transverse momentum distributions of heavy quarks whose initial transverse momenta are between 0-2, 2-5, and 5-8 GeV, respectively.
Comparing the red dashed and blue dashed dotted lines, the PHSD results have a long tail to large transverse momenta which is not present in the results of the LB equation.
For  low final $p_T$ the final distributions for low initial transverse momenta, where most of the charm quarks are located, are rather similar. This explains  the larger momentum gain and the larger variance of the transverse momentum change in PHSD as compared to LB at low initial transverse momentum, as shown in the two upper panels.

It is interesting to see that the two black lines, the sum of all contributions, are similar for both calculations, except at very low transverse momentum. Therefore we observe
a similar $R_{\rm AA}$ as shown in figure~\ref{raacD}.
We can understand this as follows: A larger drag coefficient of charm quarks  in PHSD suppresses the number of charm quarks at large transverse momentum,
but a larger diffusion coefficient compensates this suppression by spreading charm quarks from low to  large momenta. Though the momentum diffusion coefficient is of higher order than the momentum drag coefficient, it has a considerable effect for the distribution at  high momenta, because most charm quarks have initially a low $p_T$.
Although only a few charm quarks are shifted to large $p_T$  by momentum diffusion, their contribution could therefore be significant.

\begin{figure}[h!]
\centerline{
\includegraphics[width=8.6 cm]{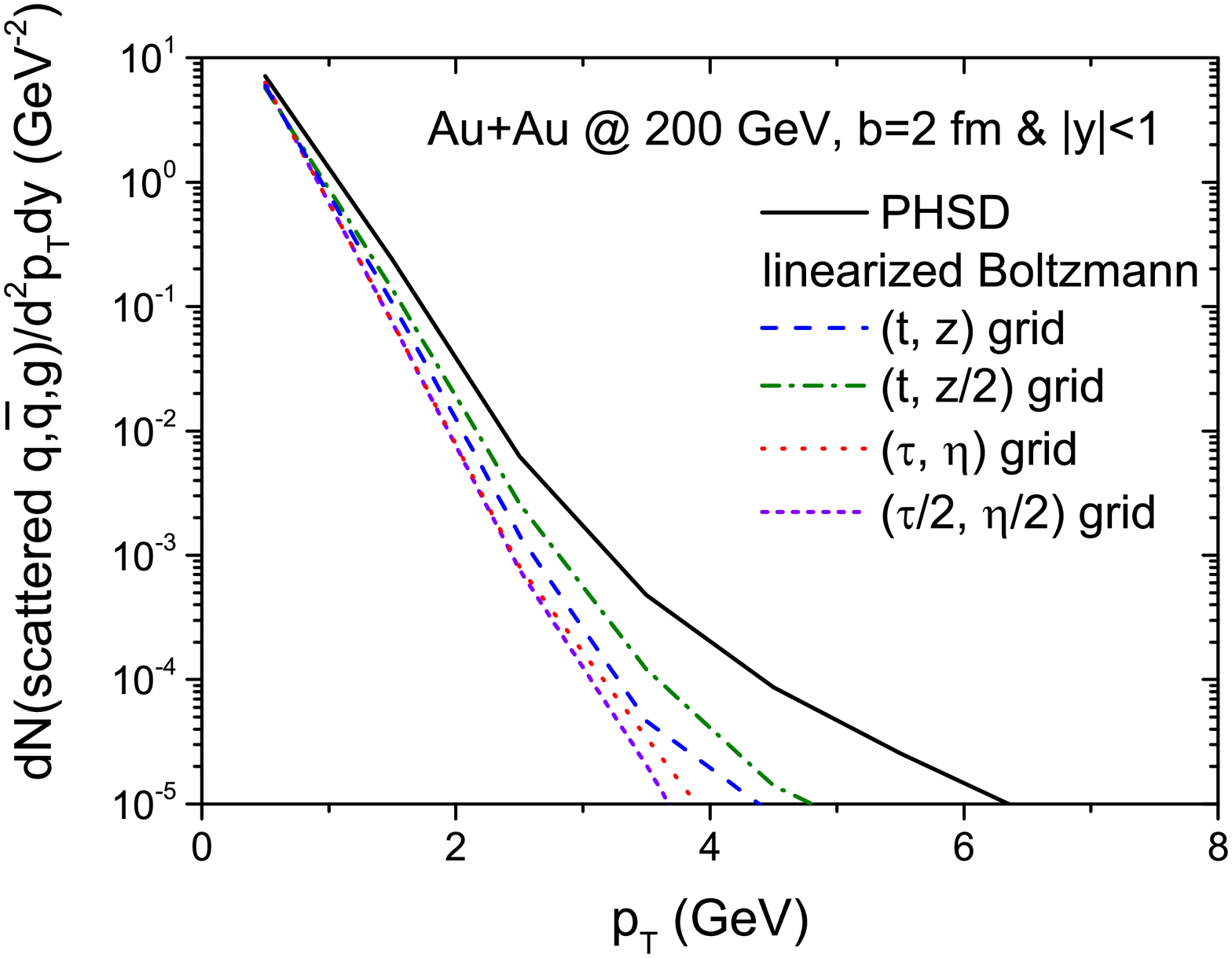}}
\caption{Transverse momentum distribution of partons which scatter off (anti)charm quarks in mid-rapidity ($|y|<1$) from the PHSD and from the linearized Boltzmann approaches.}
\label{spec_q}
\end{figure}

The reason for this large $p_T$ change of the charm quark in  PHSD calculations is elucidated  in figure~\ref{spec_q}. It shows  for both,  PHSD and LB,  the $p_T$ distribution of partons which have scattered with a (anti)charm quark, which is finally seen at mid-rapidity ($|y|<1$).
One sees that the parton spectrum in  PHSD is harder than that in the LB approach, which assumes that the cell in which the heavy quark is located is equilibrated and therefore the partons have an equilibrium distribution. This distribution is characterized by a temperature which is obtained from the energy density by the equation-of-state.
Obviously in PHSD the light partons do not have an equilibrium distribution in $p_T$ but show a strong high momentum component.  This high $p_T$ partons are responsible for the high momentum transfer observed in PHSD calculations and seen in  Fig.~\ref{microscopic} top left. This difference is large compared to the differences due to different reference frames or different grid sizes in the LB approach.  Comparing $(t,dz)$ and $(t,dz/2)$, for example, energy densities are slightly lower while transverse flow velocities are larger in $(t,dz/2)$.

We note from Fig.~\ref{spec_q} that the integral over the $p_T$ spectrum is largest in the PHSD. This means that in PHSD more collisions take place.
This is related to the increase of the cross section between heavy and light partons as a function of $\sqrt{s}$ but also to the  medium modifications of the parton mass and the parton kinetic energy in PHSD, which have been studied by some of us~\cite{Song:2019cqz} and which we explain now.

In PHSD energetic hadron scattering produces strings. If the local temperature or energy density is above the critical value for the phase transition to the QGP,
strings do not fragment into hadrons but melt into partons. This melting  is not carried out directly but through an intermediate step: in a first step hadrons, which are supposed to be produced through string fragmentation, are produced and then in a second step the hadrons are converted to partons conserving all quantum numbers as well as energy and momentum. The problem of this procedure is that in relativistic heavy-ion collisions at RHIC or LHC energies strings normally melt at very high temperatures where, according to the DQPM, on which the PHSD is based,  the partons are very massive. Therefore it may happen that the mass of the hadron which should be converted to partons is not large enough to create these massive partons. For this reason pions do not directly convert to a quark - antiquark pair  but  form first a rho meson and then  the rho meson melts into a quark-antiquark pair. Considering that a nucleon, which is composed of three constituent quarks, has a mass of around 1 GeV and a rho meson has a mass of around 0.8 GeV while the pole mass of the quark spectral function is around 0.48 GeV at 2 $T_c$, the quarks produced through the string melting have normally a mass below the pole mass in order to conserve energy and momentum. In other words, the QGP in the PHSD approach is composed of lighter quarks and antiquarks than that in the LB approach where partons are distributed according to the complete spectral distribution based on the DQPM. According to our recent study on transport coefficients of heavy quarks in non-equilibrium matter~\cite{Song:2019cqz}, heavy quarks have a larger drag and diffusion coefficient if the QGP is composed of lighter partons or whose partons have less kinetic energy than in equilibrium, assuming that the local energy density is kept constant. These results add to the explanation of the larger drag seen in PHSD calculations of figure~\ref{microscopic}.

\subsection{Forward/backward-rapidity}

The comparison between PHSD and the LB approach can be extended to forward and backward rapidities.
Presently most studies on heavy flavor production  in heavy-ion collisions are focused  on mid-rapidity, but in the future we expect
also results for forward and backward rapidities. Assuming boost invariance, the results will not depend on rapidity, but
boost invariance is only a very crude approximation. In reality it begins to break down at few rapidity units away from midrapidity.

\begin{figure}[h!]
\centerline{
\includegraphics[width=8.6 cm]{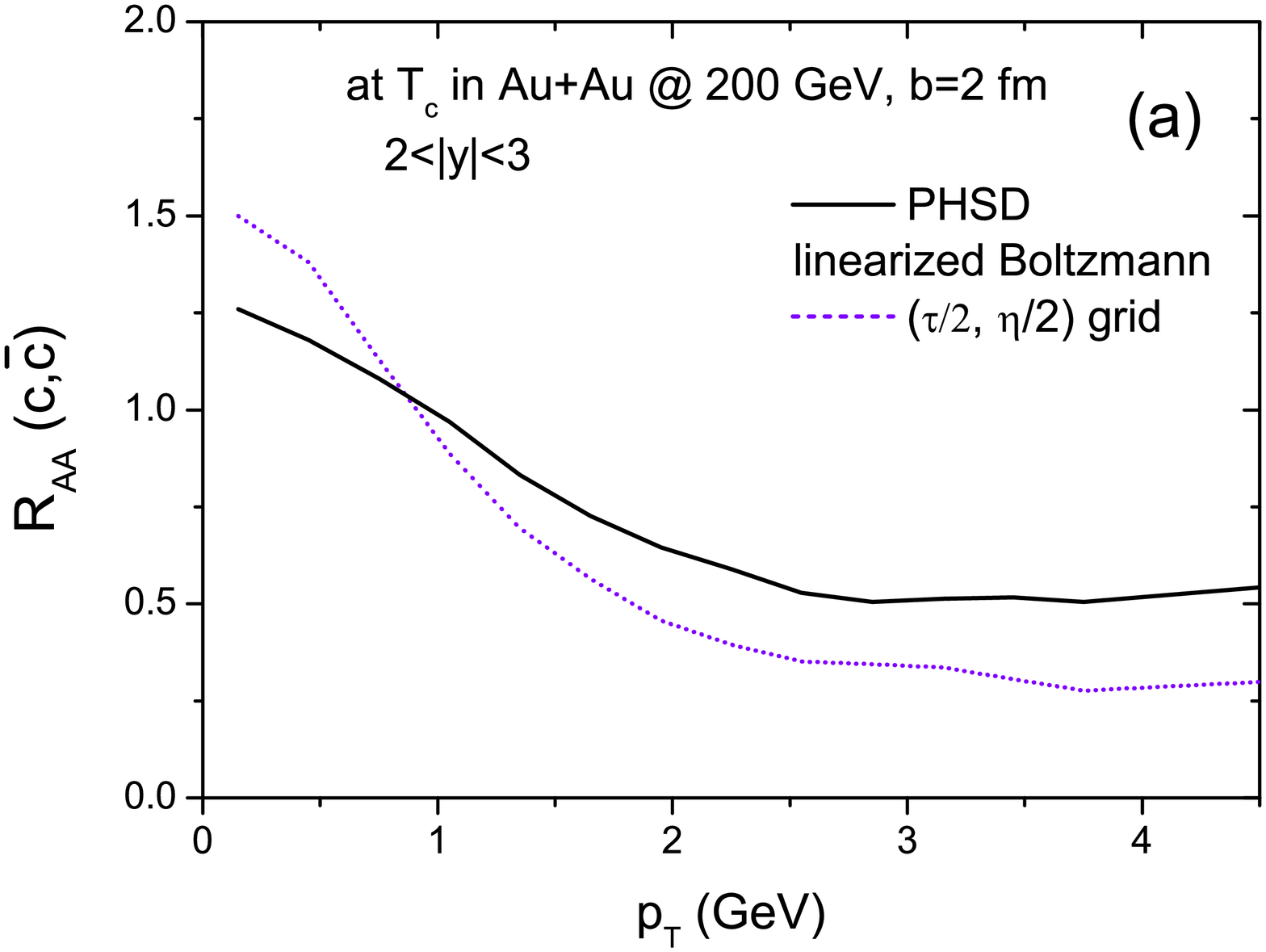}}
\centerline{
\includegraphics[width=8.6 cm]{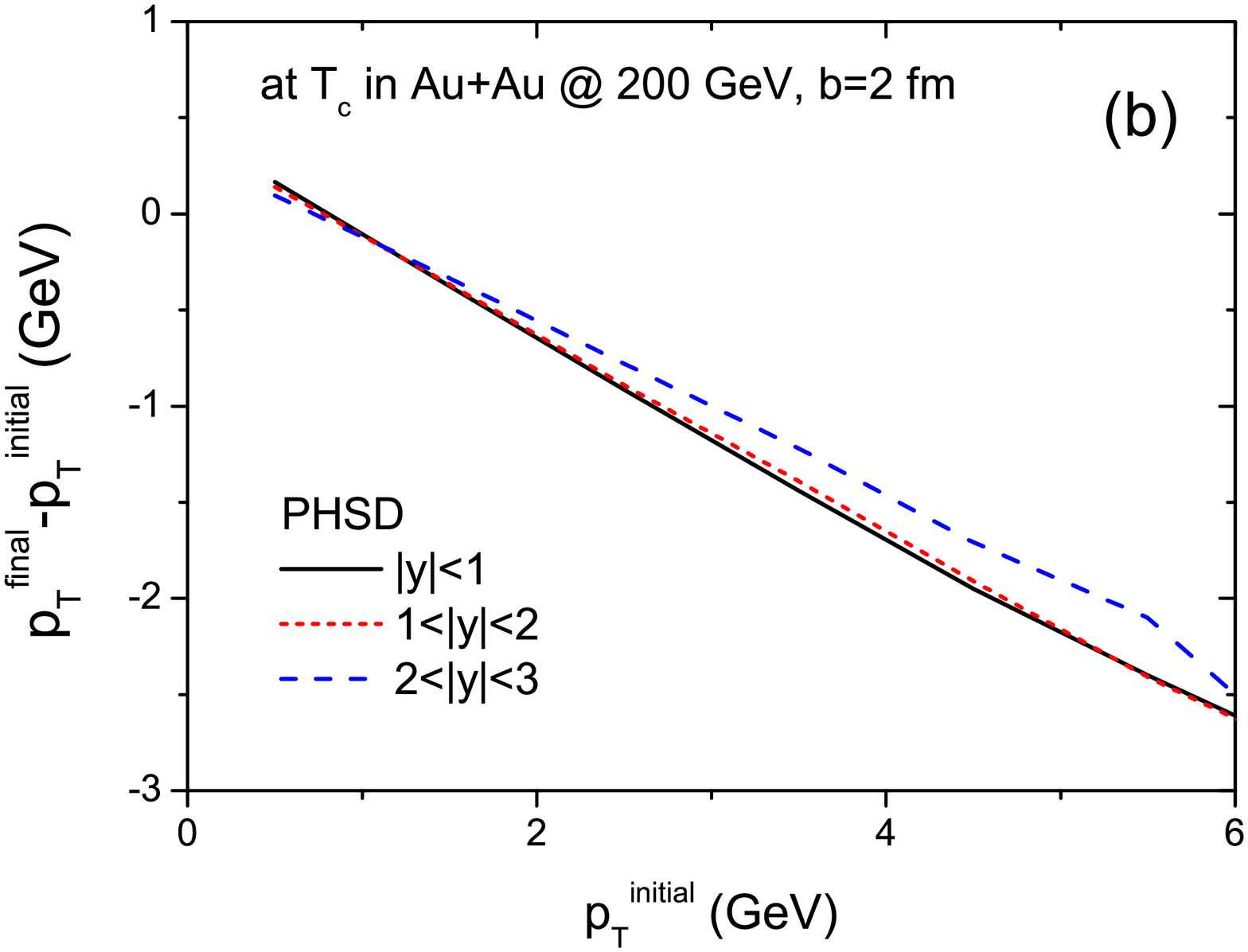}}
\centerline{
\includegraphics[width=8.6 cm]{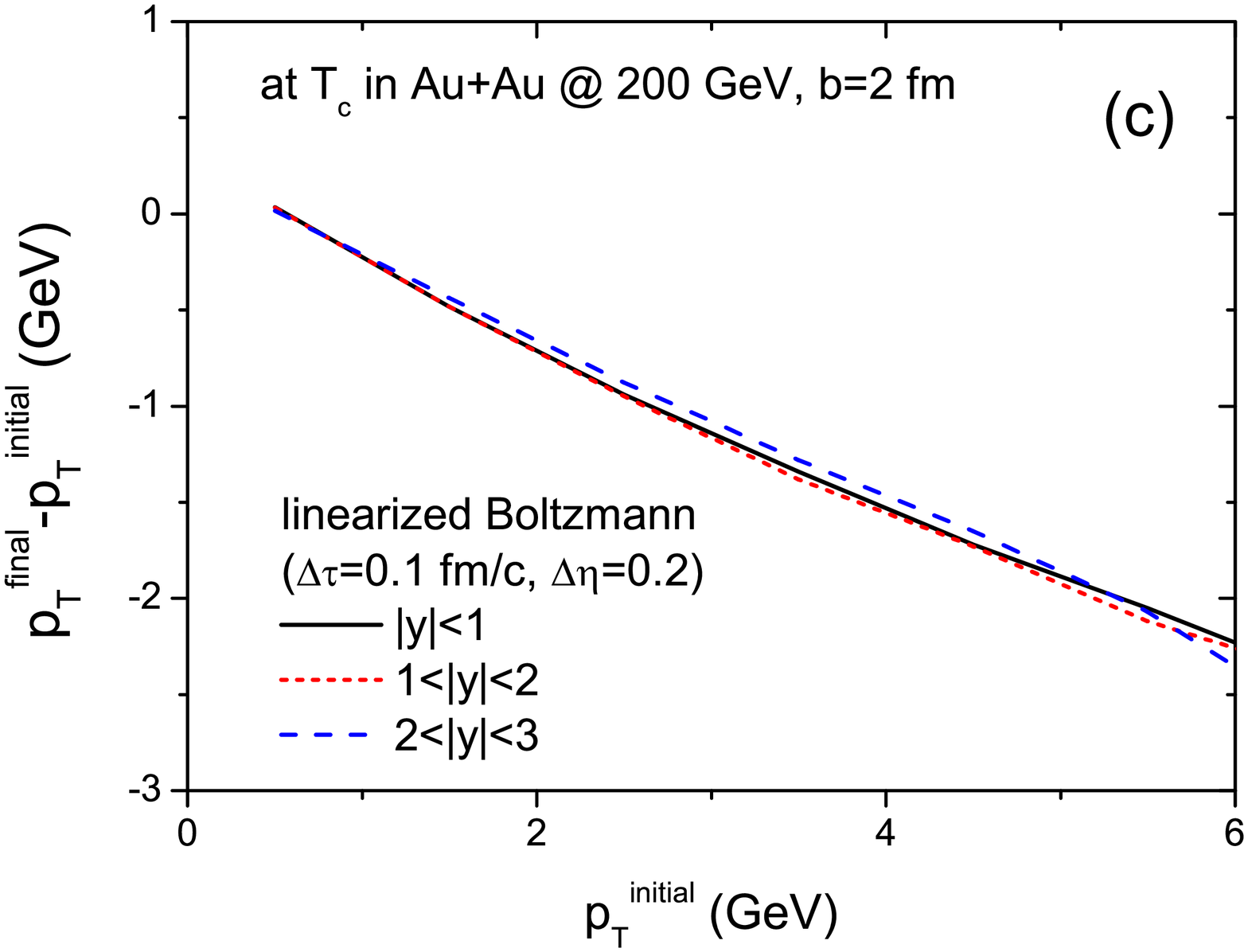}}
\caption{(a) $R_{\rm AA}(p_T)$ of (anti)charm quarks at forward/backward rapidities ($2<|y|<3$) at $T_c$ before hadronization, (b) Transverse momentum change of (anti)charm quarks in a QGP produced in Au+Au collisions at $\sqrt{s_{\rm NN}}=$ 200 GeV and $b=$ 2 fm. We display the PHSD results  as a function of the initial transverse momentum of the charm quarks and for three different rapidity ranges, and (c) same as the middle panel but for the linearized Boltzmann approach.}
\label{forward}
\end{figure}

The upper panel of figure~\ref{forward} shows $R_{\rm AA}(p_T)$ of (anti)charm quarks in forward/backward rapidities ($2<|y|<3$) at $T_c$ before hadronization in Au+Au collisions at $\sqrt{s_{\rm NN}}=$ 200 GeV and $b=$ 2 fm. Since the resolution of a grid in $(t,z)$ coordinates is not good for forward/backward rapidities, we choose for the LB approach  a grid in the $(\tau,\eta)$ coordinate system with a cell size of ($\Delta \tau=$0.1 fm/c and $\Delta \eta=$0.2).  In contrast to the results at midrapidity,
at forward rapidity the results for $R_{\rm AA}(p_T)$ differ considerably between  PHSD and the LB approach. $R_{\rm AA}(p_T)$ of charm quarks is larger at large transverse momentum in  PHSD as compared to that for the LB approach.

The middle and lower panels of figure~\ref{forward} show for a couple of rapidity bins the transverse momentum change of charm quarks as a function of their initial transverse momentum in a QGP produced in Au+Au collisions in PHSD and in the LB approach, respectively.
One finds that  in both approaches boost invariance in terms of the rapidity independence of the change of $p_T$  of charm quarks, is well satisfied up to $1<|y|<2$ . For larger
rapidities the invariance  begins to break down in PHSD, while it is still valid for the LB approach.
Comparing the middle and lower panel, we see that up to $1<|y|<2$ the drag coefficient of charm quarks is larger in PHSD than in the LB approach. In the rapidity interval $2<|y|<3$, it becomes similar in both approaches. The  larger $R_{\rm AA}(p_T)$ of charm quarks in PHSD, shown in the upper panel of figure~\ref{forward}, is due to the larger variance of the  transverse momentum change. This means that the momentum diffusion is larger which allows more charm quarks to contribute to $R_{\rm AA}$ at large transverse momentum, although in both approaches the momentum drag, as seen in middle and bottom panels, becomes similar in $2<|y|<3$.

\section{comparison with hydrodynamics}\label{yingru}

Viscous hydrodynamics, often coupled with a hadronic Boltzmann evolution for the late reaction stages, has been remarkably successful in describing the bulk evolution of ultra-relativistic heavy-ion collisions~\cite{Kolb:2000fha,Petersen:2008dd,Schenke:2010nt}. Key components of hydrodynamic calculations include initial conditions that need to be calculated with a separate initial condition model~\cite{Werner:2010aa,Schenke:2012wb}, the QCD equation of state, commonly taken from Lattice calculations~\cite{Bazavov:2014pvz,Borsanyi:2013bia,Gunther:2017sxn} and the QGP transport coefficients, most often extracted from a comprehensive model-to-data comparison~\cite{Bernhard:2016tnd,Bernhard:2019bmu}. Generally, hydrodynamics is valid under the assumption of local thermal equilibrium, even though recent kinetic theory derivations have shown the validity of hydrodynamic calculations to extend beyond that limit \cite{Romatschke:2017vte,Denicol:2018pak}.

 In contrast, PHSD provides a microscopic description of the QGP dynamics without any equilibrium assumptions. However, it does reproduce the equation-of-state and several other thermal quantities from lattice QCD in the equilibrium limit~\cite{Ozvenchuk:2012kh,Moreau:2019vhw}. The shear and bulk viscosities inherent in the PHSD dynamics can be extracted and parameterized for use in hydrodynamic calculations, making it very interesting to compare these two different dynamical approaches for the same heavy-ion collision scenario.

In a recent paper~\cite{Xu:2017pna} such a comparison has been started. It was discovered that the physics during the initial thermalization time,  before  hydrodynamic can be applied, is the critical difference between  viscous hydrodynamics and PHSD.
If PHSD and hydrodynamic simulations start with the same macroscopic initial conditions, i.e. with the temperature and the flow velocity profiles after the initial thermalization time extracted from PHSD, the results become quite similar although  PHSD displays larger fluctuations.
The ensemble averaged spatial and momentum eccentricities in PHSD are similar to those in hydrodynamics for semi-central heavy-ion collisions.
It has also been found that the initial transverse flow at the initial thermalization time has considerable effects on the dynamics of the QGP while the initial shear tensor, the off-diagonal part of energy momentum tensor, has little effect.

Many heavy flavour studies use hydrodynamics to describe the time evolution of the QGP as the underlying medium for LB calculations. Hydrodynamics provides the energy density and the flow of the grid cell  in which the heavy quark is located.
The local energy density and flow velocity of the cell are here not obtained by projecting the PHSD partons on cells, and hence by a coarse-graining of the PHSD time-evolution, but by the hydrodynamical time-evolution for a given initial condition. Consequently, comparing PHSD with hydrodynamics we can study the difference between a hydrodynamical expansion of the QGP in comparison with the PHSD dynamics. To make this possible we determine the momentum of the scattering partner of the heavy quark assuming that this momentum follows a thermal distribution in the rest frame of the cell determined from the energy density.
 Once the momentum of the QGP partons is determined we boost it from the moving cell into the center of mass of the scattering partners. The elementary collision between the heavy quark and the light parton  are described
by the Boltzmann collision integral.

While it is relatively easy to describe how heavy quarks interact with
partons from a thermalized QGP,  the heavy quark interactions with pre-equilibrium partons are not well understood. In PHSD,  partons that are produced through string melting, need a formation time, which is given by $E/m_T^2$ with $E$ and $m_T$ being energy and transverse mass, respectively. The formation time for heavy quarks is much shorter than that for light partons. During the formation time, light partons exist in form of color fields. Since it is not clear how these color fields turn into particles and how heavy quarks interact with the fields before the actual parton is formed, in PHSD it is simply assumed that the heavy quarks, after their formation time, interact with the color fields in the same way as with partons which will appear after their formation time.
On the other hand, typical hydrodynamic simulations do not extend to the pre-equilibrium stage. Because of this reason many hydrodynamical studies ignore heavy quark
interactions with partons prior to the initial thermalization time, assuming that they are negligible.
We shall therefore study first the consequences of the interaction of heavy quarks with the  initial non-equilibrium matter  before comparing  PHSD and hydrodynamics.

\begin{figure}[h!]
\centerline{
\includegraphics[width=8.6 cm]{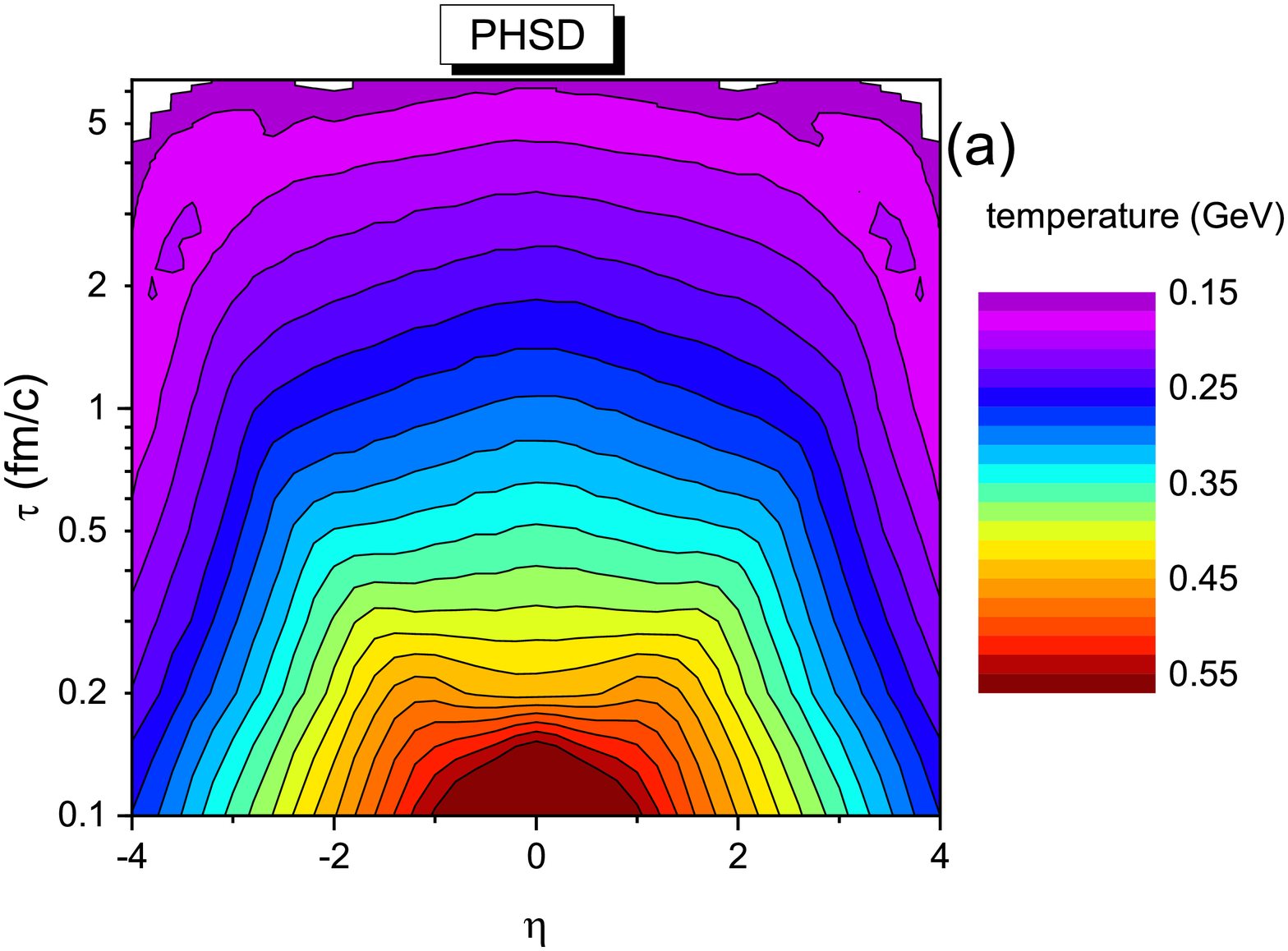}}
\centerline{
\includegraphics[width=8.6 cm]{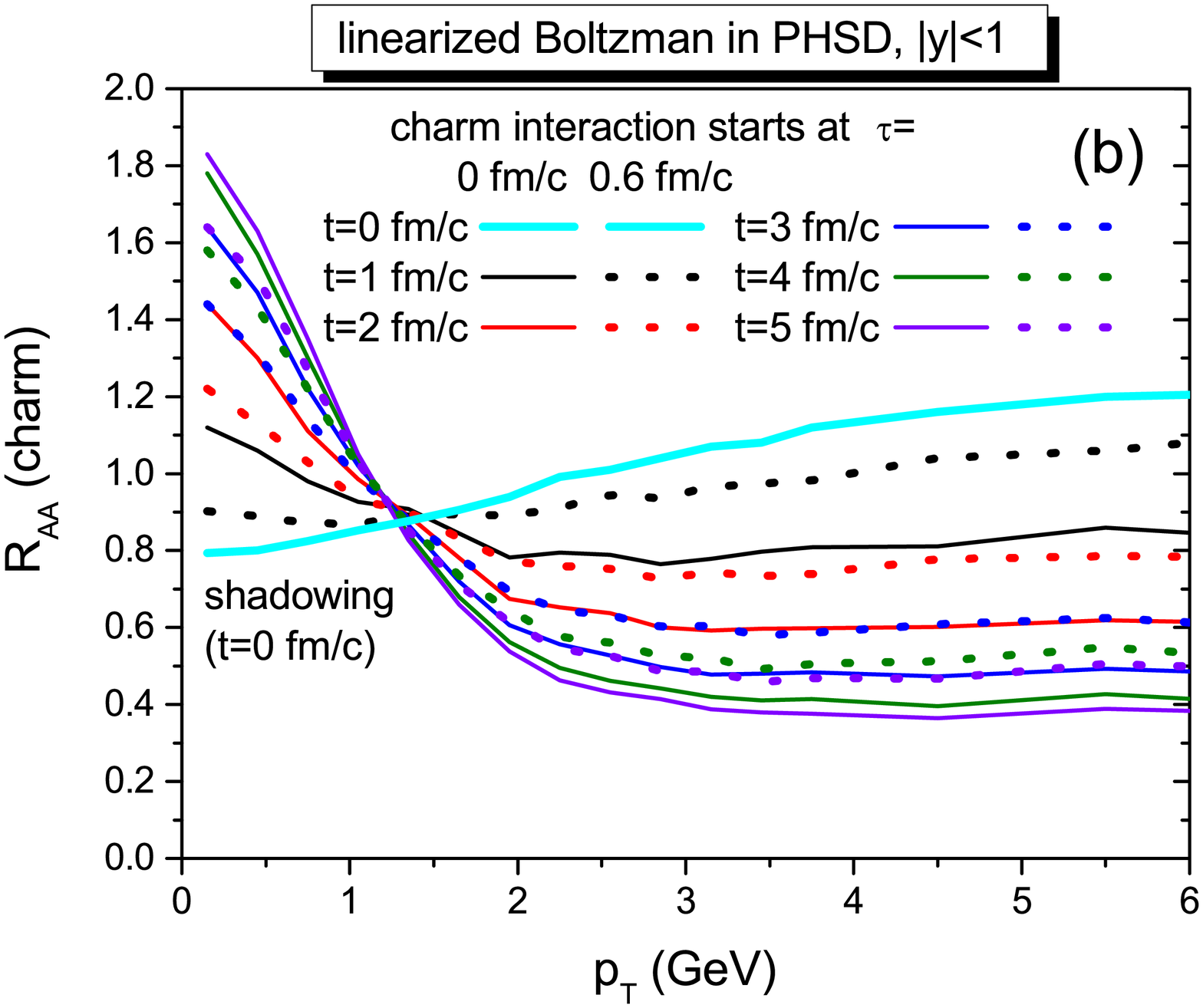}}
\caption{(a) temperature of central cells $(x=0,~ y=0)$ as a function of $\tau$ and $\eta$ in central Au+Au collisions at $\sqrt{s_{\rm NN}}=$ 200 GeV from the PHSD and (b) $R_{\rm AA}$ of charm quark in mid-rapidity ($|y|<1$) with and without charm quark interaction before $\tau=$ 0.6 fm/c are compared with each other at several time steps in linearized Boltzmann approach.}
\label{initial}
\end{figure}

The upper panel of figure~\ref{initial} shows the temperature of the central cell $(x=0,~ y=0)$ as a function of $\tau$ and $\eta$ in Au+Au collisions at $\sqrt{s_{\rm NN}}=$ 200 GeV and $b=$ 2 fm employing PHSD. The cell size is given by $\Delta \tau=0.1$ and $\Delta\eta=0.2$ in the ($\tau,\eta$) coordinate system.
One can see that boost invariance is only slightly broken at mid-rapidity.

The lower panel displays $R_{\rm AA}(p_T)$ of charm quarks at mid-rapidity ($|y|<1$) with and without charm quark interactions between their formation time and $\tau=$ 0.6 fm/c at various times during their evolution employing the LB approach. Using the EPS09 package in  PHSD~\cite{Eskola:2009uj} $R_{\rm AA}$ is already initially suppressed at low $p_T$ by shadowing effects and enhanced at large $p_T$  by anti-shadowing effects. Therefore $R_{\rm AA}$ deviates from 1 even  before the system starts to evolve ($t=$ 0 fm/c).
Comparing solid and dotted lines, where charm quarks interact in the pre-equilibrium phase and from $\tau=$ 0.6 fm/c on, respectively, we see that the early interactions have a big influence on the final value of  $R_{\rm AA}(p_T)$ in the rapidly expanding system.  The origin for this is the high temperature (see upper panel) and the high density of the environment probed by the heavy quarks at early times. This leads to a high collision rate and to a large energy transfer.

We are interested in the consequences of different dynamical evolutions of the QGP for charm quarks. Therefore we utilize the same initial condition for the time evolution of the plasma for both, PHSD and hydrodynamics. To realize this, we disable the charm quark interactions in PHSD prior to  $\tau=$ 0.6 fm/c. This yields  the dotted lines in figure~\ref{initial} which we compare to $R_{\rm AA}(p_T)$ from hydrodynamical calculations. Since the elementary cross sections are identical in both approaches the differences are then exclusively related to the different time evolution of the QGP in  PHSD and in the hydrodynamical approach.

\begin{figure*}[h!]
\centerline{
\includegraphics[width=8.6 cm]{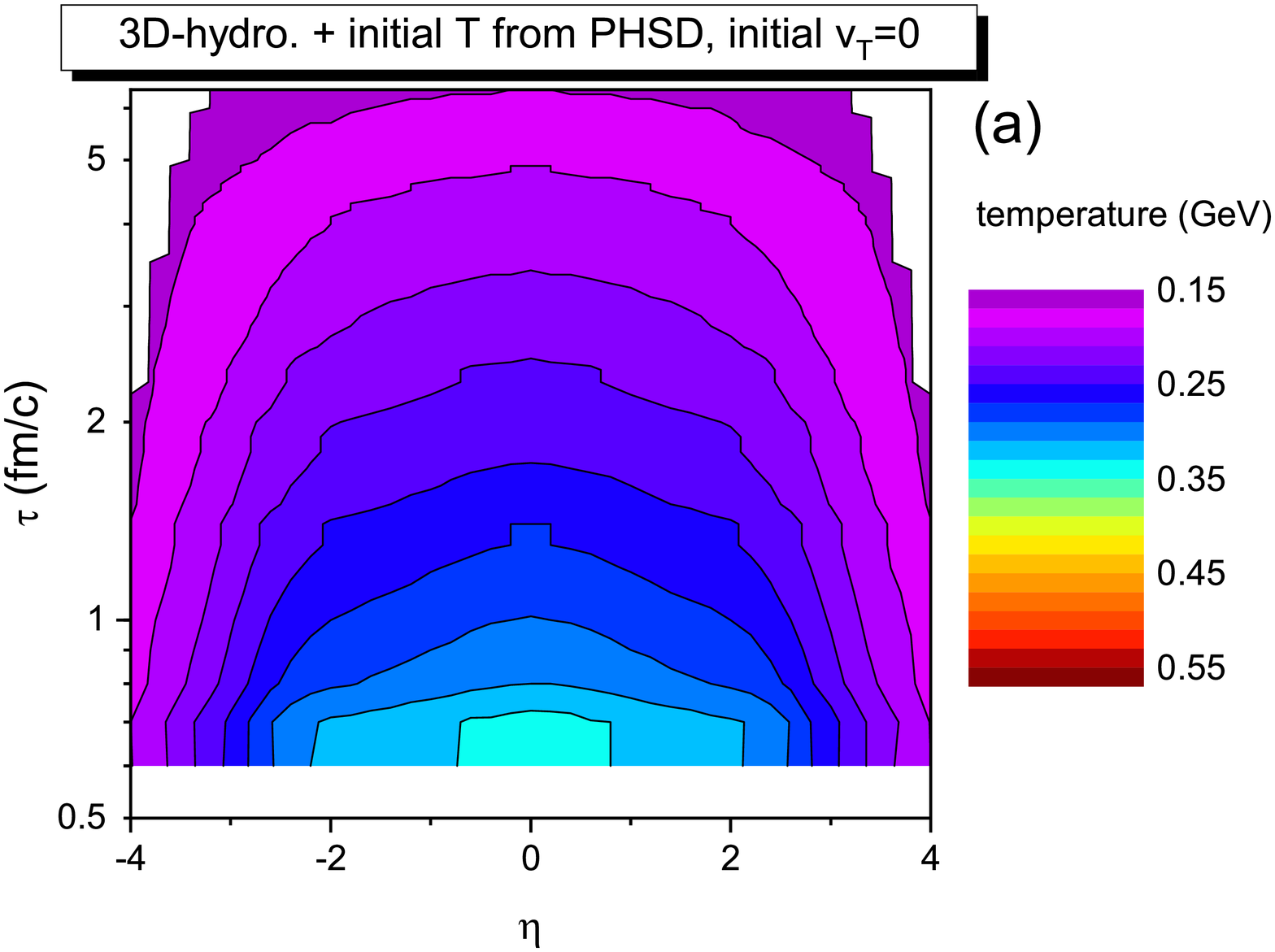}
\includegraphics[width=8.6 cm]{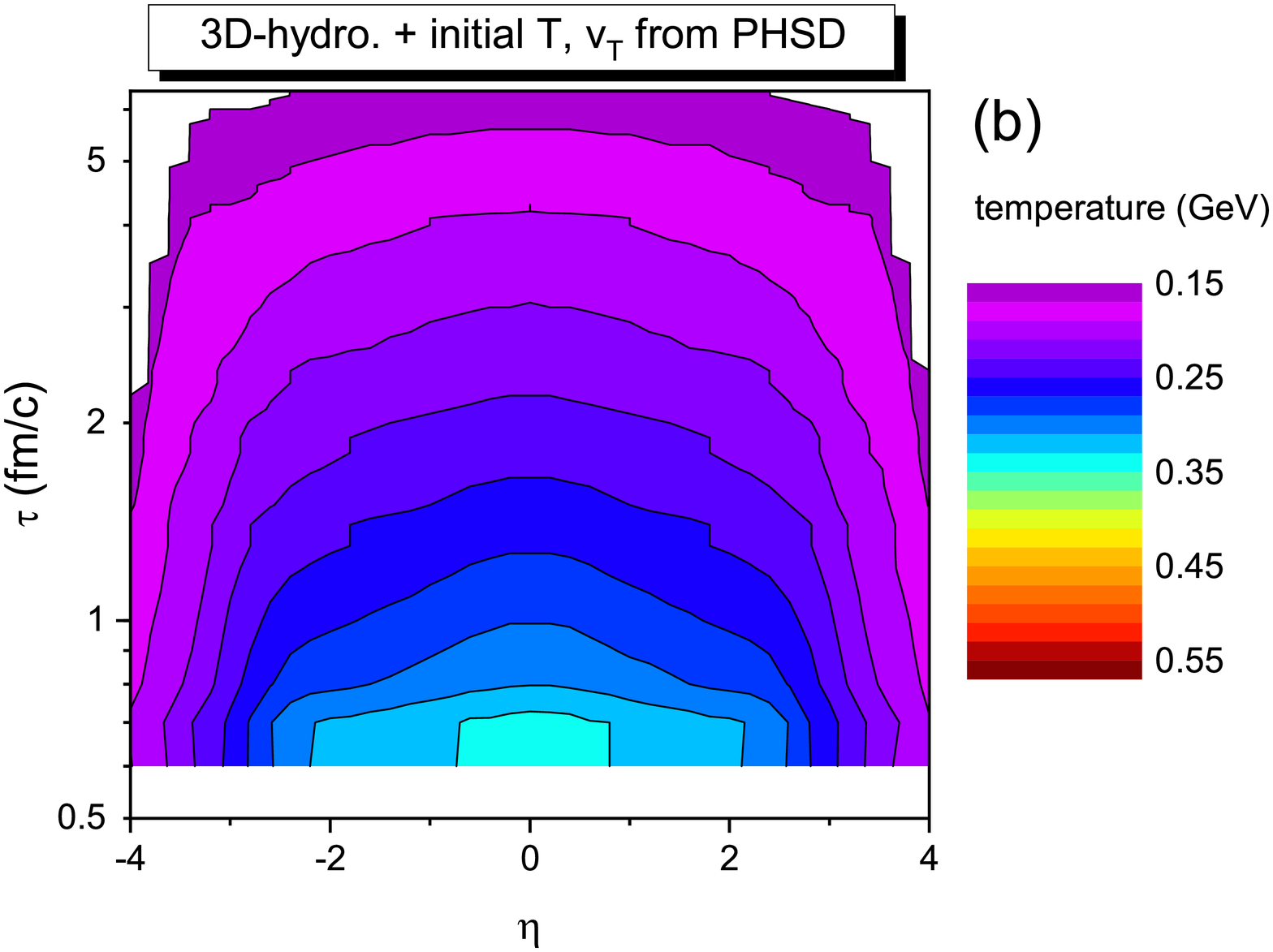}}
\centerline{
\includegraphics[width=8.6 cm]{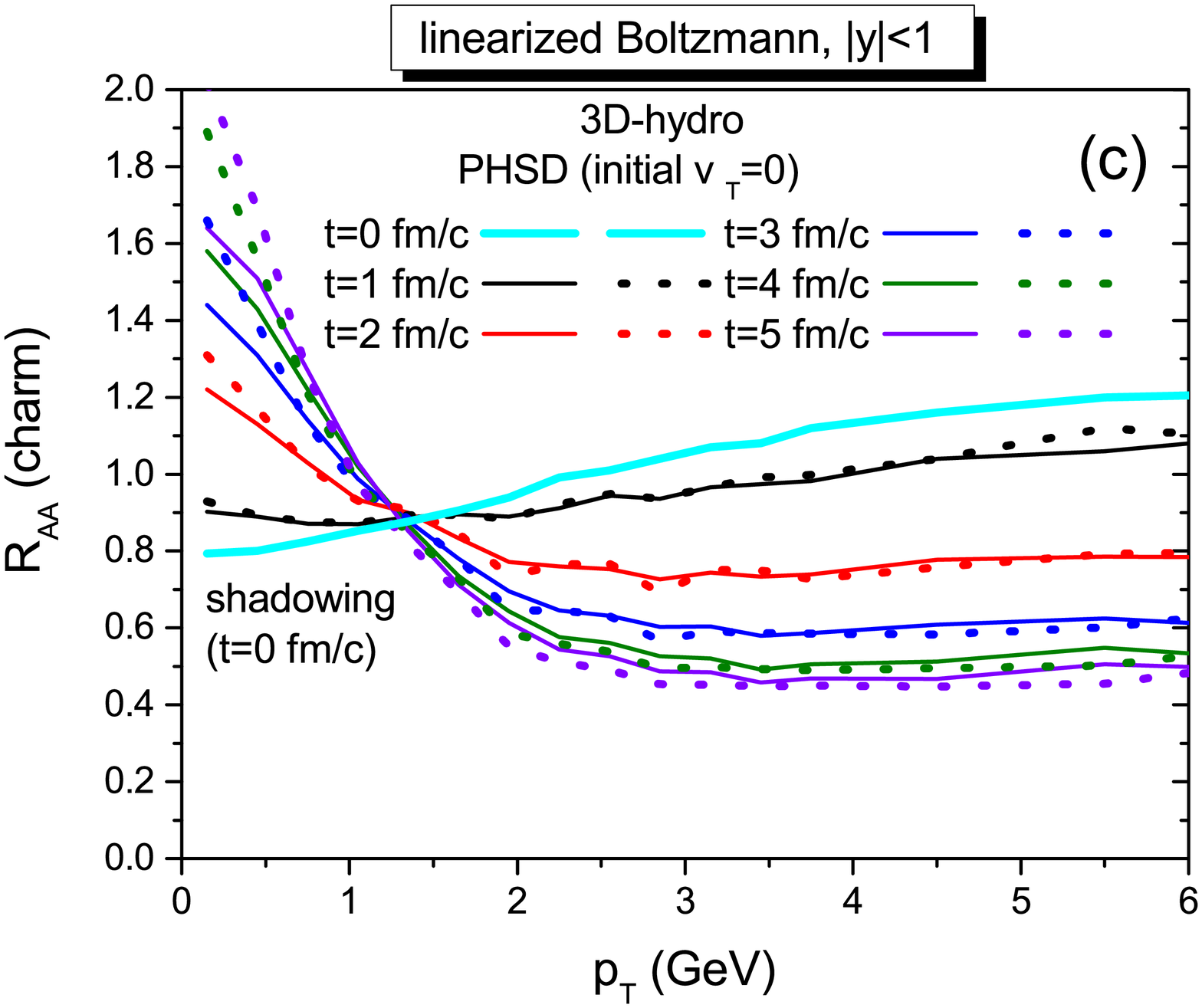}
\includegraphics[width=8.6 cm]{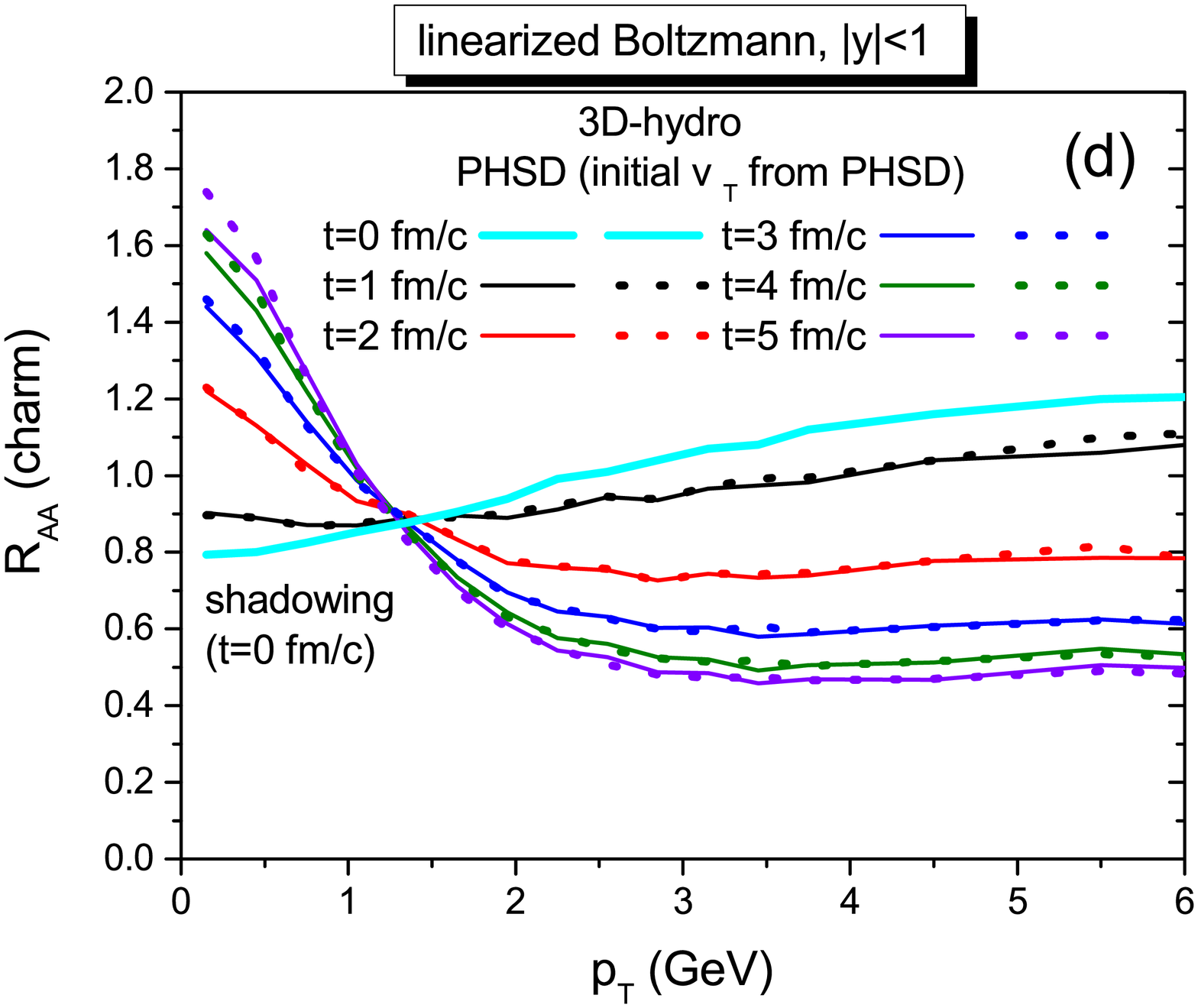}}
\caption{Temperature of the central cell $(x=0, y=0)$ as a function of $\tau$ and $\eta$ in central Au+Au collisions at $\sqrt{s_{\rm NN}}=$ 200 GeV employing 3+1 dimensional viscous hydrodynamics. In (a) we use the initial temperatures from PHSD and the longitudinal flow from boost invariance without initial transverse flow, in (b) both, initial temperatures and initial flow velocities, from PHSD. The lower panels  (c) and (d) display $R_{\rm AA}$ of mid-rapidity charm quarks  for hydrodynamical background initialized by PHSD (dotted lines) and PHSD background (full lines) in LB approach. For the calculations displayed in the left panel we assume $v_T=0$ like in the left top panel, and for those displayed in the right panel $v_T$ is given by the PHSD calculations. We assume that no charm quark - QGP interactions take place  before $\tau=$ 0.6 fm/c .}
\label{yingru-fig}
\end{figure*}

Figure~\ref{yingru-fig} shows the results from 3+1 dimensional viscous hydrodynamical calculations using the initial condition from the PHSD at $\tau=$ 0.6 fm/c in Au+Au collisions at $\sqrt{s_{\rm NN}}=$ 200 GeV and $b=$ 2 fm. Since hydrodynamics cannot be applied prior to the initial thermalization time, as discussed above, the temperature profiles in the upper panels are empty prior to $\tau=$ 0.6 fm/c. In the left panels the initial longitudinal flow is given by boost invariance and there is no initial transverse flow:

\begin{eqnarray}
v_z(\tau=0.6~{\rm fm/c},\eta)&=&\frac{z}{t}=\tanh(\eta),\nonumber\\
v_T(\tau=0.6~{\rm fm/c},\eta)&=&0.
\end{eqnarray}

In the right panels, the initial longitudinal and transverse flow velocities, as provided by  PHSD, are used in the evolution.
The calculation of the energy-momentum tensor $T^{\mu\nu}$ in the  ($\tau,\eta)$ coordinate system from the energy density and the flow velocity is given by~\cite{Karpenko:2013wva}
\begin{eqnarray}
{\rm T}^{\mu\nu}=(e+p)u^\mu u^\nu-pg^{\mu\nu},
\end{eqnarray}
where
\begin{eqnarray}
u^\tau&=&u^t\cosh\eta-u^z\sinh\eta,\nonumber\\
u^\eta&=&-u^t\sinh\eta+u^z\cosh\eta,\nonumber\\
g^{\tau\tau}&=&1,~~~g^{xx}=g^{yy}=g^{\eta\eta}=-1.
\end{eqnarray}
The initial shear tensor is ignored, because its contribution to dynamics is not significant~\cite{Xu:2017pna}.

Since there is no initial transverse flow in the left panels, the QGP cools down more slowly, which can be seen from the comparison of the upper left and right panels.
As a result, $R_{\rm AA}$ of charm quarks is  slightly lower in the left panel than in the right panel, since the lifetime of the QGP is a bit longer in the left panel.
It is interesting to note that the $R_{\rm AA}$ values in the right panel are very similar to that from the PHSD without interactions before $\tau=$ 0.6 fm/c, while the $R_{\rm AA}$ values in the left panel are slightly lower than those from PHSD.

From these comparisons in figure~\ref{yingru-fig} one can draw two conclusions:
First, the consequences  of an initial transverse flow velocity on the final spectra are not negligible, as already shown in ~\cite{Xu:2017pna} and
second, the time-evolution of the QGP, as tested by heavy quarks,  is very similar in  PHSD and in viscous hydrodynamical calculations provided that the initial conditions are identical.

\section{Comparison of PHSD and MC@HQ}\label{nantes}

\begin{figure}[H]
\centerline{
\includegraphics[width=8.6 cm]{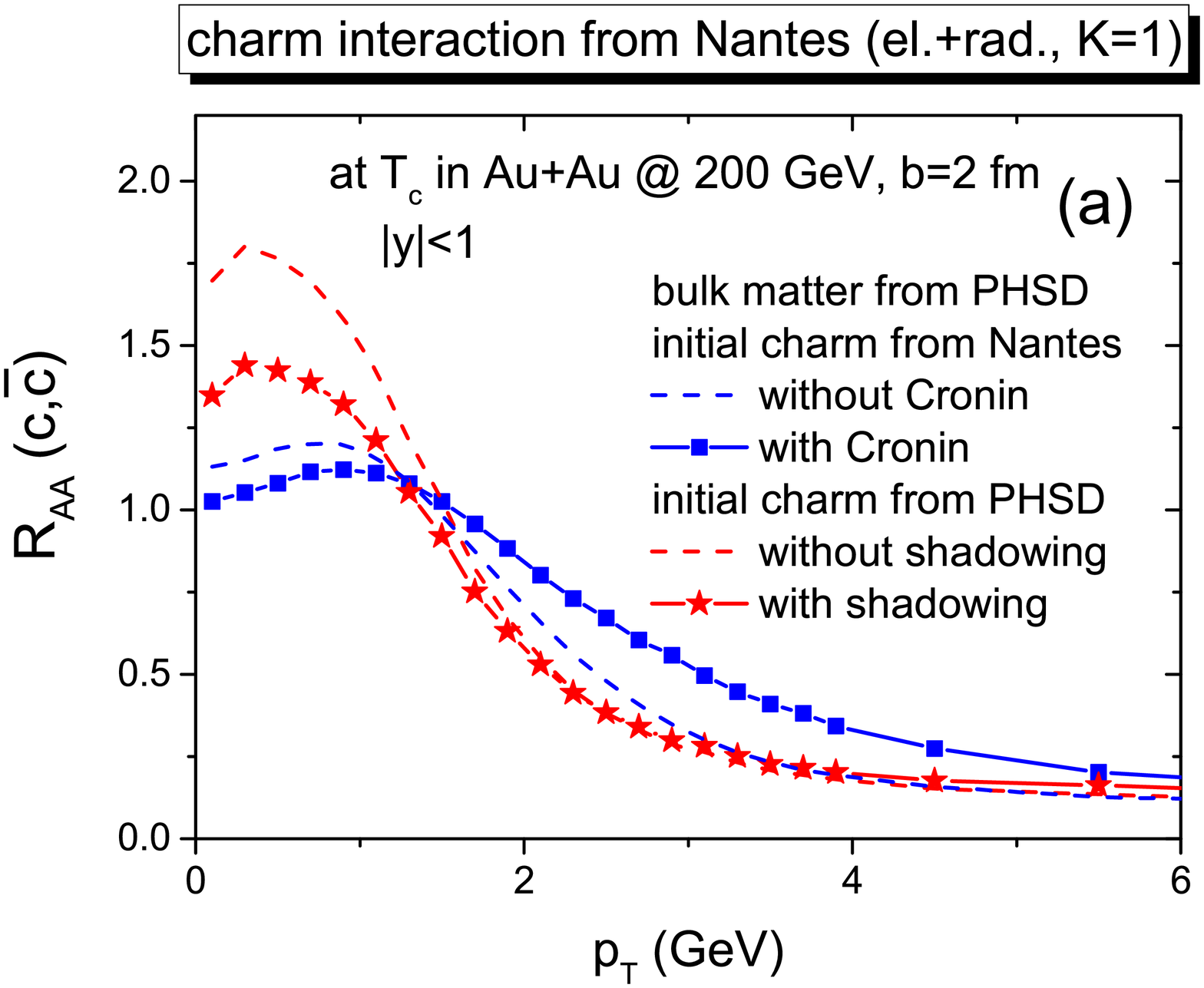}}
\centerline{
\includegraphics[width=8.6 cm]{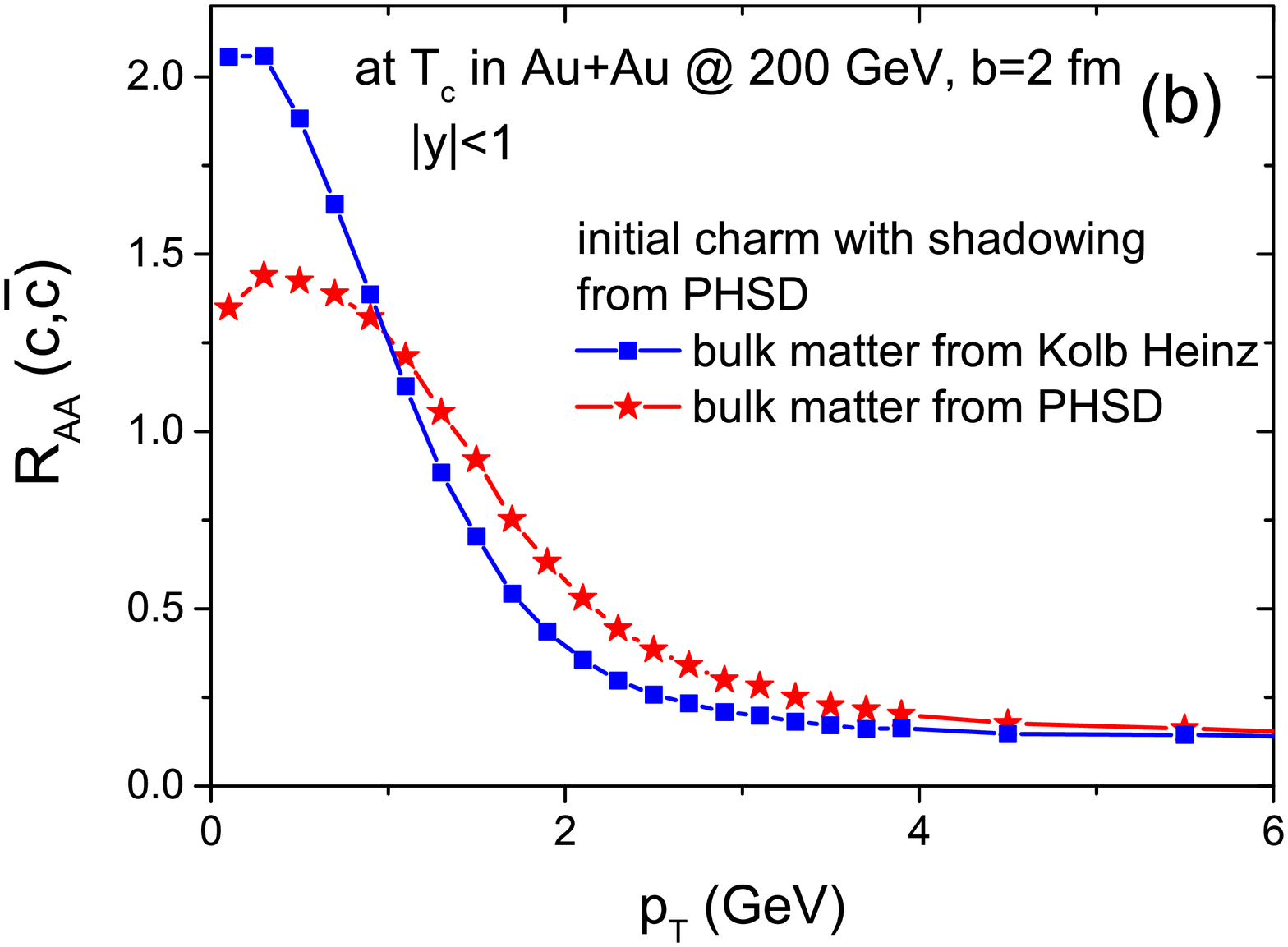}}
\centerline{
\includegraphics[width=8.6 cm]{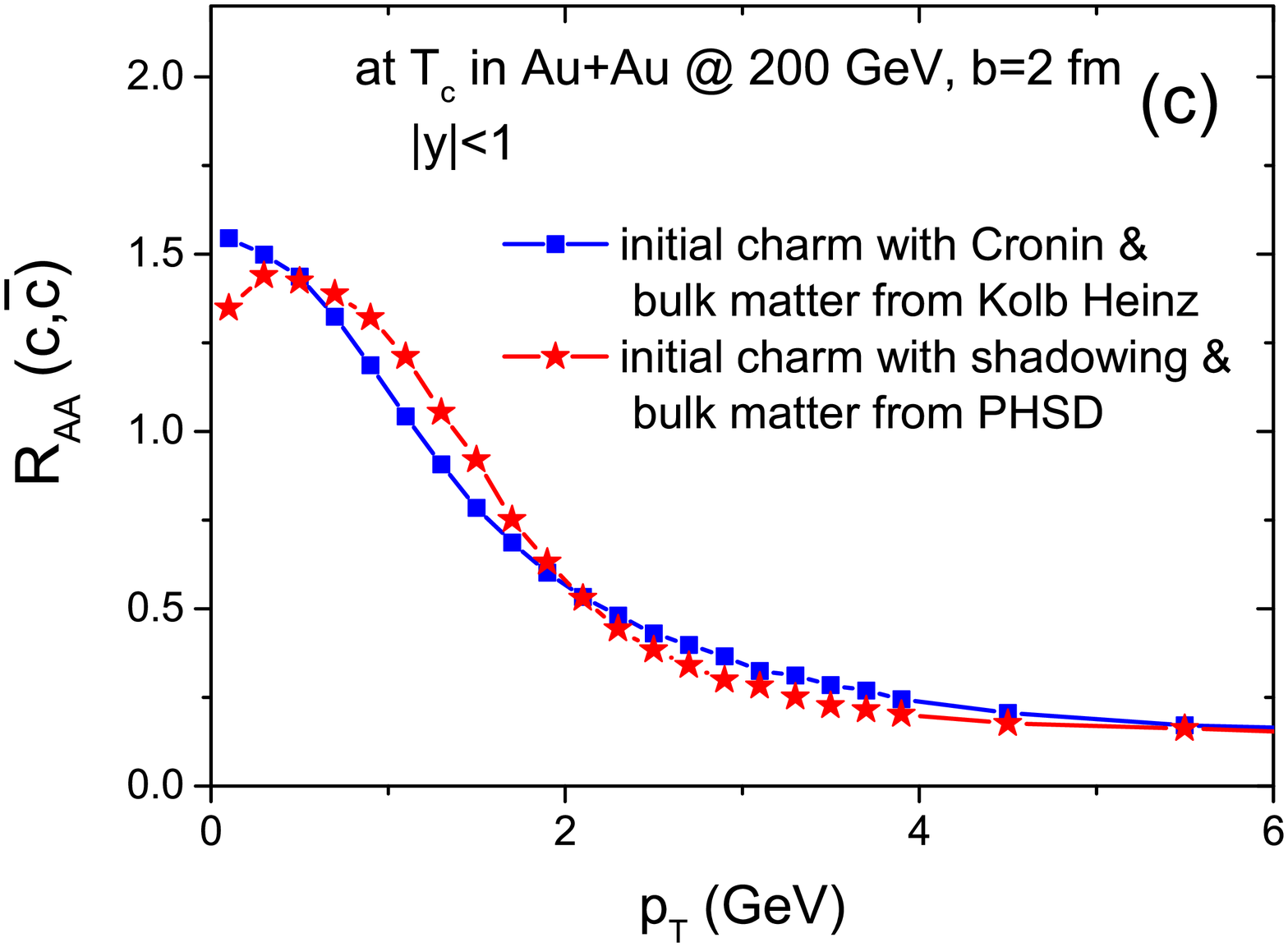}}
\caption{$R_{\rm AA}$ of mid-rapidity (anti)charm quarks at $T_c$ (before hadronization) in central Au+Au collisions at $\sqrt{s_{\rm NN}}=$ 200 GeV. We compare the influence of different initial charm spectrum and of different QGP evolutions on this observable. top:
Influence of different initial charm spectrum. We compare the results for the initial charm spectrum of the Nantes approach (with and without Cronin effect) with that for the PHSD initial charm spectrum (with and without shadowing). The QGP evolution is from PHSD. Middle:   Influence of different time evolutions of the QGP for the same (PHSD) initial charm spectrum. Bottom: Result of the standard MC@HQ approach (initial charm spectrum and QGP evolution from Kolb Heinz+MC@HQ) compared with that of standard PHSD (initial charm spectrum and QGP evolution from PHSD).
}
\label{pol}
\end{figure}

So far we have focused our study on the effects of non-equilibrium QGP on charm dynamics. 
In this section we extend the discussion to the consequences of different initial charm quark
and different elementary interactions between heavy quarks and partons from the QGP. For this we use an additional approach to study open heavy flavor observables, MC@HQ, which has been developed by the Nantes group and combined with different hydrodynamical scenarios that describe the expansion of the plasma, namely the one from Kolb Heinz for RHIC energies~\cite{Kolb:2000fha,Kolb:2004}. Both approaches use FONLL calculations for the initial charm quark spectrum.
This description is not unique as the calculation for p+p collisions shows: The spectrum of the Nantes model is close to the upper bounds of the FONLL calculations at low transverse momentum while PHSD always takes the mean values of FONLL. The elementary interaction differs in three essential points from that of the PHSD approach: The QGP partons are massless, the coupling constant depends on the momentum transfer (and not on the temperature) and the interactions between the heavy quarks and the QGP partons can also be inelastic.
The inelastic collisions are those in which a gluon is emitted in addition to the particles in the entrance channel. For details we refer to \cite{Aichelin:2013mra}.  These newly created gluons are affected by the Landau-Pomeranchuck-Migdal effect which states that they need time to be considered as independent (created) particles. This effect is taken into account in the Nantes approach \cite{Gossiaux:2012cv}. To perform each collision one picks, as in the LB of PHSD, randomly the momentum of the colliding parton (q,g) from the local thermal distribution in the hydro-cell. This parton collides with the heavy quark according to cross sections which are calculated with the lowest order Feynman diagrams. The elastic cross section differs from the pQCD cross section by having a running coupling constant ($\alpha (t)$) and a modified propagator. Instead of a propagation $\propto (t-m_D^2)^{-1}$, the form $\propto (t-\kappa m_D^2  )^{-1}$ is used where $\kappa$  is determined by the requirement that the energy loss is independent from the intermediate scale which separates the HTL dominated low momentum transfer from the Born diagram which describes the cross section for high momentum transfer, following the procedure which Braaten and Thoma have introduced for QED \cite{Braaten:1991jj}.  The K-factor is taken to be 1, which means that high-order corrections are ignored in the pQCD calculations.

Figure~\ref{pol} compares $R_{\rm AA}$ of (anti)charm quarks observed at midrapidiy at $T_c$ (before hadronization) in central Au+Au collisions at $\sqrt{s_{\rm NN}}=$ 200 GeV. We display the influence of different initial charm spectra and of different descriptions of the expansion of the QGP. The interaction between the charm quarks and the QGP follows the Nantes model.

In the upper panel we study the influence of different initial charm quark spectra on $R_{\rm AA}$ of charm quarks at $T_c$ (before hadronization). The expansion of the QGP is described by the PHSD.
Both, the  Nantes approach and  PHSD,  include cold nuclear matter effects, the Cronin effect in the former and shadowing effects in the latter.
The Cronin effect is the enhancement of the heavy quark transverse momentum due to the scattering of a nucleon in one nucleus and a parton of the other nucleus such that the parton gains additional transverse momentum before the hard scattering which produces heavy flavor~\cite{Hufner:2001tg}.
As expected, the Cronin effect suppresses $R_{\rm AA}$ at low transverse momentum and enhances it at large transverse momentum.
In a nucleus the number of partons  at small $x$, with $x$ being the  longitudinal momentum fraction, decreases and that at large $x$ increases.
The former is called shadowing and the latter antishadowing. The (anti)shadowing effects suppress $R_{\rm AA}$ at low transverse momentum and enhance it at large transverse momentum, as the Cronin effect. Whether the (anti)shadowing effect includes the Cronin effect or not is controversial.
We display the results for two different PHSD initial charm spectrum (with and without shadowing) and two different Nantes initial charm spectrum (with and without Cronin effect). We observe that the different initial conditions have a strong influence on $R_{\rm AA}$ at $T_c$, especially at low $p_T$. Since the Cronin effect shifts the whole $p_T$ distribution it is still visible at intermediate $p_T$ whereas the antishadowing  is only little visible. At low momentum  PHSD  shows an enhanced yield as compared to the Nantes model whereas at large charm quark momenta the approaches become more similar.

The middle panel shows how different descriptions of the expansion of the QGP, those from MC@HQ (namely, Kolb Heinz for RHIC energies) and from PHSD, influence $R_{\rm AA}$  at $T_c$ . Here
we use the grid of PHSD as it is used in the LB approach as described in sections III and IV.  Initial charm spectrum (PHSD) and elementary interactions (from Nantes) are the same for both models. It is clearly visible that the hydrodynamical expansion in the Nantes model yields a larger enhancement at small $p_T$ than the PHSD expansion.
$R_{\rm AA}$ for the Kolb Heinz expansion are below those for the PHSD expansion for $1.5< p_T<4$ GeV.

The lower panel compares the consequences from choosing standard ingredients from the PHSD approach as compared to the ones from MC@HQ. We see that the effects observed in a)  and b) compensate each other to a large extent. The higher  $R_{\rm AA}$  in PHSD due to the PHSD initial charm spectrum is compensated by the lower  $R_{\rm AA}$  due to PHSD expansion of the QGP and vice versa.

\begin{figure}[H]
	\centerline{
		\includegraphics[width=8.6 cm]{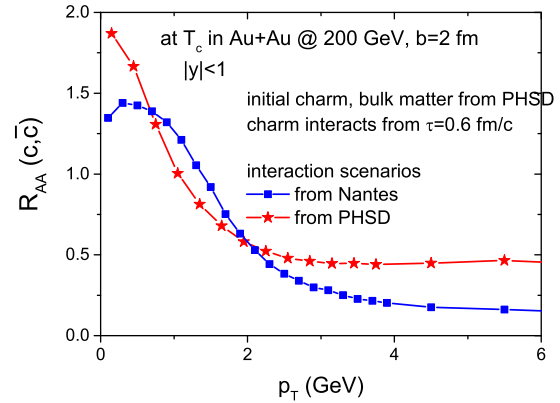}}
	\caption{$R_{\rm AA}$ of mid-rapidity (anti)charm quarks at $T_c$ (before hadronization) for the elementary interaction between  heavy quarks and partons in the QGP from EPOS+MC@HQ (blue) and  from PHSD in the LB version (red line). Charm quarks start to interact after $\tau=$ 0.6 fm/c. The initial charm spectrum and the expansion of the QGP are taken from the PHSD model.}
	\label{pol2}
\end{figure}

Besides the initial charm spectrum and the QGP expansion there is a third component which has influence on   $R_{\rm AA}$  at $T_c$, the elementary interaction between heavy quarks and QGP partons. This influence is addressed in Fig.~\ref{pol2}. It shows $R_{\rm AA}$ of mid-rapidity (anti)charm quarks at $T_c$ for different elementary interactions between heavy quarks and QGP partons. The expansion of the QGP and initial charm quarks distribution are given by the PHSD. Charm quarks start to interact after $\tau=$ 0.6 fm/c.  For the red curve (which is identical to the full red curves in Figure~\ref{pol}) the elementary  interaction is taken from the PHSD approach whereas for the blue curve
the interaction of the Nantes approach is applied. We see also here a considerable difference in  $R_{\rm AA}$.
Though it is beyond the scope of the present study, we note that the $R_{\rm AA}$ shown here is that of the heavy quark at hadronization and cannot be compared to experimental results for heavy mesons. D mesons can be created by coalescence of a QGP quark or by fragmentation. The relative fraction of both depends on $p_T$.
The coalescence probability of charm quarks is larger in MC@HQ than in PHSD. This decreases the differences of $R_{\rm AA}$, because the coalescence increases $p_T$ and,
as a result, suppresses $R_{\rm AA}$ at small $p_T$ and enhances it at large $p_T$.

Figures~\ref{pol} and \ref{pol2} show the challenges regarding the use of  charm quarks to study properties of the QGP produced in heavy-ion collisions. The lifetime of the plasma is rather short due to the fast expansion. Therefore differences in the initial state of the system show up in the final charm quark spectra prior to hadronization. Different initial charm spectra, different expansion scenarios as well as a different elementary interactions between heavy quarks and partons of the QGP lead to $p_T$ dependent modifications of  $R_{\rm AA}$ which may easily reach individually 50\% but which may compensate each other. Therefore models with  different expansions, different elementary interactions and different initial charm spectra may yield a similar final state  $R_{\rm AA}(p_T)$.   Consequently, with the available experimental data,  essentially $R_{\rm AA}$ and $v_2$ at midrapidty,  all measured with a considerable error, one cannot yet identify precisely the contributions of the different sources to the deviation of  $R_{\rm AA}(p_T)$ from unity.


\section{summary}\label{summary}

Heavy flavor mesons are one of the few probes which allow to investigate the properties of the QGP which is produced in relativistic heavy-ion collisions.
This is due to a couple of advantages. The production of heavy quarks can be reliably described by perturbative QCD as the comparison of pp data with pQCD calculations shows.
Intermediate and high $p_T$ heavy quarks do not come to an equilibrium with the expanding QGP and carry therefore information on their interaction with the QGP constituents. The interaction of heavy mesons with the hadronic environment at the end of the reaction can be calculated,
though its contribution to the nuclear modification factor of heavy meson is limited at RHIC and LHC energies. There remain uncertainties about how heavy mesons are produced at the end of the expansion of the QGP, but for high $p_T$ quarks some fragmentation function, for example Petersen fragmentation function,
can be applied. For a given initial distribution of heavy quarks the final heavy meson observables are sensitive to two unknown ingredients in the time evolution equations of heavy quarks and mesons: to the expansion of the QGP and to the elementary interaction of the heavy quarks with the QGP constituents.

In the first part of this  manuscript we have assumed that the QGP is composed of quasi-particles given by the dynamical quasi-particle model and that the elementary interactions can be described by first order Feynman diagrams between these quasi-particles and the heavy quarks. The quasi-particle properties are obtained by the requirement that in thermal equilibrium these quasi-particles yield an equation-of-state as given by the lattice QCD calculations. The PHSD approach  is not the only possibility to describe these interactions between heavy quarks and plasma constituents but it has been very successful in describing the experimental results in the light quark sector.

Subsequently, we have investigated how different descriptions of the evolution of the bulk medium affects the heavy quark observables.
Here we studied three different models for the QGP expansion.\\  A) a hydrodynamical approach which is based on the assumption that the system is  in local equilibrium during its expansion. This approach is based on conservation laws and requires as only input the equation-of-state of strongly interacting matter, which has been calculated for vanishing chemical potential by lQCD calculations. The drawback of this approach is its reliance on local thermal equilibrium and the need for an intial condition that has to be obtained prior to the hydrodynamical evolution.\\  B) The PHSD approach which assumes that the QGP is composed of quasi-particles whose time evolution is given by the Kadanoff-Baym equations.\\  C) The coarse grained PHSD approach in which a grid is introduced on which the partons, propagated by the PHSD equations, are projected. At each time step and for each cell the energy density and the cell flow velocity are calculated. The equation-of-state allows to convert the energy density into a temperature and it is assumed that the distribution of the particles in the cell is a thermal distribution with the temperature of the cell. Thus local equilibrium is assumed, however the time evolution of the local energy density is not given by the hydrodynamical time evolution equations but by that of PHSD.  Due to this assumption of local thermal equilibrium the time evolution of the heavy quarks in the QGP is given by a linearized Boltzmann equation. The grid in this approach can either be in $(t,~z)$ or in $(\tau,~\eta)$ coordinates. We have found that a grid in the $(t,~z)$ coordinate system provides a finer resolution at mid-rapidity, while a grid in $(\tau,~\eta)$ coordinate system is better suited for forward and backward rapidities.
Assuming that boost invariance is a good approximation near mid-rapidity at the top energy of RHIC, a grid in $(\tau,~\eta)$ coordinate system with the cell size of ($\Delta \tau=$ 0.1 fm/c, $\Delta \eta=$ 0.2) is used for most comparisons in this study.

Since the linearized Boltzmann approach assumes local thermal equilibrium, the comparison of the results of PHSD with that using a linearized Boltzmann approach reveals to what extent a local equilibrium is established in PHSD. We found that at mid-rapidity charm quarks lose more energy at large transverse momentum without the assumption of a local thermal equilibrium. This translates into a larger drag coefficient of charm quarks in PHSD. It shows that coarse graining of transport theories (in order to use for example microscopically calculated thermal production rates of heavy quarks or linearized Boltzmann equations)  may bias results and has to be tested against fully microscopic calculations.

Although the results from PHSD and from  LB differ in details for the central observable in heavy-ion collisions, $R_{\rm AA}(p_T)$, the influence of this larger drag coefficient is compensated by a larger diffusion coefficient. Therefore  $R_{\rm AA}(p_T)$  of charm quarks is similar at large transverse momentum independent of whether a local thermal equilibrium is assumed or not. Also $v_2$ of  heavy quarks and D mesons are rather insensitive to the assumption of a local thermal equilibrium.

Extending the comparison to forward and backward rapidities we see that the  boost invariance begins to break down earlier in the PHSD without the assumption of thermal equilibrium and that the drag coefficient of charm quark momentum starts to decrease for rapidities of $2<|y|<3$. While, assuming thermal equilibrium, the drag coefficient is nearly rapidity independent, it gets smaller in PHSD. As a result, $R_{\rm AA}$ of charm quarks in PHSD without thermal equilibrium is larger than that in the LB approach.

To test whether the space-time evolution of the energy density and of  the collective velocity is different we compared charm quark interactions in the QGP described by PHSD and by 3+1 dimensional viscous hydrodynamics with the initial conditions provided by PHSD, (both calculated in the LB approach).  This comparison shows that after  $\tau = 0.6$ fm/c
both approaches give very similar results. Consequently the elementary interaction among the partons in PHSD are sufficiently strong
for macroscopic thermal quantities to follow hydrodynamics, though the matter still remains in non-equilibrium microscopically.
This justifies a posteriori also the parametrization of the masses and coupling constant in PHSD as a function of the local temperature.

When we compare the heavy quark observables calculated in PHSD and in viscous hydrodynamics, the difference comes from the interactions between heavy quarks and their environment before $\tau = 0.6$ fm/c when the system has obtained a local equilibrium and therefore hydrodynamical calculations can start. In PHSD partons are produced through string melting and are ready for interactions after their formation time which depends on the transverse mass of the particle. Since the interactions of charm quarks with the not yet formed QGP partons is not well known, we assume that it is same as the interaction with formed partons. As a consequence, $R_{\rm AA}$ of charm quark is more suppressed by about 0.1 at large transverse momentum ($4<p_{\rm T}<6$ GeV) if charm quarks are allowed to interacts before the initial thermalization time.
Considering that $R_{\rm AA}$ of charm quarks is around 0.4 at $4<p_{\rm T}<6$ GeV, the effect is not negligible.

We have also found that the initial transverse flow, which is sometimes neglected in hydrodynamic simulations, has an effect on charm quark observables, though this influence is not as strong  as that from the interactions before the initial thermalization. If the initial transverse flow is ignored, the cooling of the QGP becomes a bit slower and charm quarks interact in QGP for a longer time. As a result, $R_{\rm AA}$ of charm quarks is slightly lower than in the PHSD calculations.

In the second part of this article we studied the influence of the elementary cross section between heavy quarks and QGP partons on $R_{\rm AA}(p_T)$. We compared  for two approaches, the Nantes and the PHSD approach, those quantities which influence $R_{\rm AA}$ of heavy quarks at $T_c$, before they hadronize (to eliminate the uncertainties due to different hadronization models and due to hadronic final state interactions). For this purpose we modified the three ingredients of kinetic approaches, the heavy quark initial distribution, the QGP expansion and the elementary interaction between heavy quarks and QGP partons, independently, keeping the other two ingredients fixed. We see that in all three cases the modification of  $R_{\rm AA}$  is  not negligible and $p_T$ dependent. Models, in which all three ingredients are rather different, may nevertheless give very similar  $R_{\rm AA}$  values, as has been observed in the past~\cite{Gossiaux:2011ea,Andronic:2015wma,Rapp:2018qla,Cao:2018ews,Xu:2018gux}. Therefore, the observables at hand will not allow to unambiguously determine these ingredients separately. One may hope that with new experimental data models like EPOS+MC@HQ and PHSD, which describe not only heavy quarks but also the light quark observables, can be used to limit the uncertainties of the QGP expansion and that heavy-ion reactions with different size nuclei as well as correlations between heavy mesons may constrain the elementary interaction between heavy and light quarks further. At this stage, it will for sure be mandatory for each model to take the off-equilibrium effects into account, as they are one component of possible discrepancies, however not dominant over the other ones we have investigated in this work.

\section*{Acknowledgements}

This work was supported by the LOEWE center "HIC for FAIR", the HGS-HIRe for FAIR,
the COST Action THOR, CA15213, and
the German Academic Exchange Service (DAAD) (T.S.,
P.M., E.B.) and by the
National Science Centre, Poland under Grant No. 2014/14/E/ST2/00018
(V.O.).
Furthermore, PM and EB acknowledge support by DFG through the grant CRC-TR 211 'Strong-interaction matter under extreme conditions'.
Major computational resources have been provided by the LOEWE-CSC.This project has received funding from the European Union’s Horizon 2020 research and innovation program under grant agreement No 824093 (Strong-2020). Y.X. and S.A.B. have been supported by the U.S Department of Energy under grant DE-FG02-05ER41367.
J.A., M.N. and P.B.G. have been supported by Region Pays de la Loire (France) under contract no. 2015-08473.
M.N. acknowledges the support by the Region Pays de la Loire, France, under a "Etoiles montantes" grant, and in part by the ExtreMe Matter Institute EMMI at the
GSI Helmholtzzentrum für Schwerionenforschung, Darmstadt, Germany.


\begin{thebibliography}{99}

\bibitem{Cacciari:1998it}
  M.~Cacciari, M.~Greco and P.~Nason,
  JHEP {\bf 9805}, 007 (1998).

\bibitem{Cacciari:2001td}
  M.~Cacciari, S.~Frixione and P.~Nason,
  JHEP {\bf 0103}, 006 (2001).

\bibitem{Vogt}
M. Cacciari, P. Nason and R. Vogt, Phys. Rev. Lett. {\bf 95},  122001 (2005).



\bibitem{Moore:2004tg}
  G.~D.~Moore and D.~Teaney,
  Phys.\ Rev.\ C {\bf 71}, 064904 (2005).

\bibitem{vanHees:2005wb}
  H.~van Hees, V.~Greco and R.~Rapp,
  Phys.\ Rev.\ C {\bf 73}, 034913 (2006).

\bibitem{Uphoff:2012gb}
  J.~Uphoff, O.~Fochler, Z.~Xu and C.~Greiner,
  Phys.\ Lett.\ B {\bf 717}, 430 (2012).


\bibitem{He:2014cla}
  M.~He, R.~J.~Fries and R.~Rapp,
  Phys.\ Lett.\ B {\bf 735}, 445 (2014).

\bibitem{Cao:2013ita}
  S.~Cao, G.~Y.~Qin and S.~A.~Bass,
  Phys.\ Rev.\ C {\bf 88}, 044907 (2013).

\bibitem{Gossiaux:2010yx}
  P.~B.~Gossiaux, J.~Aichelin, T.~Gousset and V.~Guiho,
  J.\ Phys.\ G {\bf 37}, 094019 (2010).

\bibitem{Das:2015ana}
  S.~K.~Das, F.~Scardina, S.~Plumari and V.~Greco,
  Phys.\ Lett.\ B {\bf 747}, 260 (2015).

\bibitem{Song:2015sfa}
  T.~Song, H.~Berrehrah, D.~Cabrera, J.~M.~Torres-Rincon, L.~Tolos, W.~Cassing and E.~Bratkovskaya,
  Phys.\ Rev.\ C {\bf 92}, no. 1, 014910 (2015).

\bibitem{Song:2015ykw}
  T.~Song, H.~Berrehrah, D.~Cabrera, W.~Cassing and E.~Bratkovskaya,
  Phys.\ Rev.\ C {\bf 93}, no. 3, 034906 (2016).

\bibitem{Xu:2017obm}
  Y.~Xu, J.~E.~Bernhard, S.~A.~Bass, M.~Nahrgang and S.~Cao,
  Phys.\ Rev.\ C {\bf 97}, no. 1, 014907 (2018).

\bibitem{Ke:2018tsh}
  W.~Ke, Y.~Xu and S.~A.~Bass,
  Phys.\ Rev.\ C {\bf 98}, no. 6, 064901 (2018).

\bibitem{Svetitsky:1987gq}
  B.~Svetitsky,
  Phys.\ Rev.\ D {\bf 37}, 2484 (1988).



\bibitem{Song:2019cqz}
  T.~Song, P.~Moreau, J.~Aichelin and E.~Bratkovskaya,
  arXiv:1910.09889 [nucl-th].

\bibitem{Cassing:2009vt}
  W.~Cassing and E.~L.~Bratkovskaya,
  Nucl.\ Phys.\ A {\bf 831}, 215 (2009).


\bibitem{PHSD}
W. Cassing and E.L. Bratkovskaya, Nucl. Phys. A {\bf 831}, 215 (2009).

\bibitem{PHSDrhic}
  E.~L.~Bratkovskaya, W.~Cassing, V.~P.~Konchakovski and O.~Linnyk,
  Nucl.\ Phys.\ A {\bf 856}, 162 (2011).

\bibitem{Volo} V. P. Konchakovski {\it et al.}, J. Phys. G {\bf 42}, 055106  (2015);  J. Phys. G {\bf 41}, 105004  (2014); Phys. Rev. C {\bf 90}, 014903 (2014);
Phys. Rev. C {\bf 85},  044922 (2012);  Phys. Rev. C {\bf 85}, 011902 (2012).

\bibitem{Linnyk} O. Linnyk {\it et al.},  Phys. Rev. C {\bf 89},   034908 (2014);  Phys. Rev. C {\bf 88},  034904 (2013); Phys. Rev. C {\bf 87},   014905 (2013); Phys. Rev. C {\bf 85},  024910 (2012); Phys. Rev. C {\bf 84}, 054917 (2011); Nucl. Phys. A {\bf 855}, 273 (2011).







\bibitem{Sjostrand:2006za}
  T.~Sjostrand, S.~Mrenna and P.~Z.~Skands,
  JHEP {\bf 0605}, 026 (2006).

\bibitem{Eskola:2009uj}
  K.~J.~Eskola, H.~Paukkunen and C.~A.~Salgado,
  JHEP {\bf 0904}, 065 (2009).

\bibitem{Berrehrah:2013mua}
H.~Berrehrah, E.~Bratkovskaya, W.~Cassing, P.~B.~Gossiaux, J.~Aichelin and M.~Bleicher,
Phys.\ Rev.\ C {\bf 89}, no. 5, 054901 (2014).

\bibitem{Moreau:2019vhw}
P.~Moreau, O.~Soloveva, L.~Oliva, T.~Song, W.~Cassing and E.~Bratkovskaya,
Phys.\ Rev.\ C {\bf 100}, no. 1, 014911 (2019).



\bibitem{Song:2016rzw}
  T.~Song, H.~Berrehrah, J.~M.~Torres-Rincon, L.~Tolos, D.~Cabrera, W.~Cassing and E.~Bratkovskaya,
  Phys.\ Rev.\ C {\bf 96}, no. 1, 014905 (2017).

\bibitem{Song:2018xca}
  T.~Song, W.~Cassing, P.~Moreau and E.~Bratkovskaya,
  Phys.\ Rev.\ C {\bf 97}, no. 6, 064907 (2018).

\bibitem{Song:2018dvf}
  T.~Song, W.~Cassing, P.~Moreau and E.~Bratkovskaya,
  Phys.\ Rev.\ C {\bf 98}, no. 4, 041901 (2018).


\bibitem{Xu:2017pna}
  Y.~Xu, P.~Moreau, T.~Song, M.~Nahrgang, S.~A.~Bass and E.~Bratkovskaya,
  Phys.\ Rev.\ C {\bf 96}, no. 2, 024902 (2017).



\bibitem{Gossiaux:2009mk}
  P.~B.~Gossiaux, R.~Bierkandt and J.~Aichelin,
  Phys.\ Rev.\ C {\bf 79}, 044906 (2009).






\bibitem{Werner:2010aa}
  K.~Werner, I.~Karpenko, T.~Pierog, M.~Bleicher and K.~Mikhailov,
  Phys.\ Rev.\ C {\bf 82}, 044904 (2010).

\bibitem{Bass:1998ca}
  S.~A.~Bass {\it et al.},
  Prog.\ Part.\ Nucl.\ Phys.\  {\bf 41}, 255 (1998).



\bibitem{Bleicher:1999xi}
  M.~Bleicher {\it et al.},
  J.\ Phys.\ G {\bf 25} (1999) 1859
  [hep-ph/9909407].

\bibitem{Aichelin:2013mra}
  J.~Aichelin, P.~B.~Gossiaux and T.~Gousset,
  Phys.\ Rev.\ D {\bf 89}, no. 7, 074018 (2014).

\bibitem{Nahrgang:2013saa}
  M.~Nahrgang, J.~Aichelin, P.~B.~Gossiaux and K.~Werner,
  Phys.\ Rev.\ C {\bf 90}, no. 2, 024907 (2014).


\bibitem{Nahrgang:2014vza}
  M.~Nahrgang, J.~Aichelin, S.~Bass, P.~B.~Gossiaux and K.~Werner,
  Phys.\ Rev.\ C {\bf 91}, no. 1, 014904 (2015).

\bibitem{Nahrgang:2013xaa}
  M.~Nahrgang, J.~Aichelin, P.~B.~Gossiaux and K.~Werner,
  Phys.\ Rev.\ C {\bf 89}, no. 1, 014905 (2014).

\bibitem{Linnyk:2015rco}
O.~Linnyk, E.~L.~Bratkovskaya and W.~Cassing,
Prog.\ Part.\ Nucl.\ Phys.\  {\bf 87}, 50 (2016).






\bibitem{Adamczyk:2014uip}
  L.~Adamczyk {\it et al.} [STAR Collaboration],
  Phys.\ Rev.\ Lett.\  {\bf 113}, no. 14, 142301 (2014)
  Erratum: [Phys.\ Rev.\ Lett.\  {\bf 121}, no. 22, 229901 (2018)].

\bibitem{Lomnitz:2016rpz}
  M.~R.~Lomnitz [STAR Collaboration],
  Nucl.\ Phys.\ A {\bf 956}, 256 (2016).


\bibitem{Schenke:2010nt}
  B.~Schenke, S.~Jeon and C.~Gale,
  Phys.\ Rev.\ C {\bf 82}, 014903 (2010).

\bibitem{Kolb:2000fha}
  P.~F.~Kolb, P.~Huovinen, U.~W.~Heinz and H.~Heiselberg,
  Phys.\ Lett.\ B {\bf 500}, 232 (2001).

\bibitem{Petersen:2008dd}
  H.~Petersen, J.~Steinheimer, G.~Burau, M.~Bleicher and H.~Stocker,
  Phys.\ Rev.\ C {\bf 78}, 044901 (2008).








\bibitem{Schenke:2012wb}
  B.~Schenke, P.~Tribedy and R.~Venugopalan,
  Phys.\ Rev.\ Lett.\  {\bf 108}, 252301 (2012).

\bibitem{Bazavov:2014pvz}
  A.~Bazavov {\it et al.} [HotQCD Collaboration],
  Phys.\ Rev.\ D {\bf 90}, 094503 (2014).

\bibitem{Borsanyi:2013bia}
  S.~Borsanyi, Z.~Fodor, C.~Hoelbling, S.~D.~Katz, S.~Krieg and K.~K.~Szabo,
  Phys.\ Lett.\ B {\bf 730}, 99 (2014).

\bibitem{Gunther:2017sxn}
  J.~Günther, R.~Bellwied, S.~Borsanyi, Z.~Fodor, S.~D.~Katz, A.~Pasztor and C.~Ratti,
  EPJ Web Conf.\  {\bf 137}, 07008 (2017).

\bibitem{Bernhard:2016tnd}
  J.~E.~Bernhard, J.~S.~Moreland, S.~A.~Bass, J.~Liu and U.~Heinz,
  Phys.\ Rev.\ C {\bf 94}, no. 2, 024907 (2016).

\bibitem{Bernhard:2019bmu}
  J.~E.~Bernhard, J.~S.~Moreland and S.~A.~Bass,
  Nature Phys.\  {\bf 15}, no. 11, 1113 (2019).

\bibitem{Romatschke:2017vte}
  P.~Romatschke,
  Phys.\ Rev.\ Lett.\  {\bf 120}, no. 1, 012301 (2018)
  doi:10.1103/PhysRevLett.120.012301
  [arXiv:1704.08699 [hep-th]].

\bibitem{Denicol:2018pak}
  G.~S.~Denicol and J.~Noronha,
  Phys.\ Rev.\ D {\bf 99}, no. 11, 116004 (2019)
  doi:10.1103/PhysRevD.99.116004
  [arXiv:1804.04771 [nucl-th]].

\bibitem{Ozvenchuk:2012kh}
  V.~Ozvenchuk, O.~Linnyk, M.~I.~Gorenstein, E.~L.~Bratkovskaya and W.~Cassing,
  Phys.\ Rev.\ C {\bf 87}, no. 6, 064903 (2013).





\bibitem{Karpenko:2013wva}
  I.~Karpenko, P.~Huovinen and M.~Bleicher,
  Comput.\ Phys.\ Commun.\  {\bf 185}, 3016 (2014).

\bibitem{Kolb:2004}
P. Kolb and U. Heinz, in Quark Gluon Plasma 3, edited by
R. Hwa and X. N. Wang (World Scientific, Singapore, 2004),
p. 634.



\bibitem{Gossiaux:2012cv}
P.B.~Gossiaux
Nuclear Physics, Section A 910-911, 301 (2013).

\bibitem{Braaten:1991jj}
  E.~Braaten and M.~H.~Thoma,
  Phys.\ Rev.\ D {\bf 44}, 1298 (1991).

\bibitem{Hufner:2001tg}
  J.~Hufner and P.~f.~Zhuang,
  Phys.\ Lett.\ B {\bf 515}, 115 (2001).


\bibitem{Gossiaux:2011ea}
  P.~B.~Gossiaux, S.~Vogel, H.~van Hees, J.~Aichelin, R.~Rapp, M.~He and M.~Bluhm,
  [arXiv:1102.1114 [hep-ph]].

\bibitem{Andronic:2015wma}
  A.~Andronic {\it et al.},
  Eur.\ Phys.\ J.\ C {\bf 76}, no. 3, 107 (2016).

\bibitem{Rapp:2018qla}
  R.~Rapp {\it et al.},
  Nucl.\ Phys.\ A {\bf 979}, 21 (2018).

\bibitem{Cao:2018ews}
  S.~Cao {\it et al.},
  Phys.\ Rev.\ C {\bf 99}, no. 5, 054907 (2019).

\bibitem{Xu:2018gux}
  Y.~Xu {\it et al.},
  Phys.\ Rev.\ C {\bf 99}, no. 1, 014902 (2019).











\end{thebibliography}
\end{document}